\documentclass[aps,twocolumn]{revtex4}
\usepackage{graphics}
\usepackage{graphicx}
\usepackage{bm}
\usepackage{latexsym}
\usepackage{color}
\usepackage{mathrsfs}
\usepackage{braket}
\usepackage{enumerate}
\usepackage[usenames,dvipsnames]{xcolor}
\usepackage[colorlinks=true,citecolor=blue,linkcolor=blue,urlcolor=blue,backref=true]{hyperref}
\usepackage{float}
\usepackage[raggedright,tight]{subfigure}  %side-by-side figures
\usepackage{amsmath}
\usepackage{amsfonts}
\usepackage{braket}
\usepackage[normalem]{ulem}
\usepackage{cancel}
\usepackage{mathtools}
\usepackage{appendix}
\usepackage{upgreek}
\usepackage{accents}
\usepackage{tikz}
\usepackage{cleveref}
\usetikzlibrary{positioning}
\begin{document}

\title{Regularised Arbitrary Gauge non-Relativistic QED}
\author{Alex Chivers-White}\email{a.white13@newcastle.com}
\author{Adam Stokes}
\affiliation{School of Mathematics, Statistics, and Physics, Newcastle University, Newcastle upon Tyne, NE1 7RU, United Kingdom}

\begin{abstract}
%We develop a fully regularised arbitrary-gauge formulation of non-relativistic QED for short-range light–matter interactions and apply it to two stationary hydrogen atoms. The theory yields formally distinct but unitarily equivalent gauge-fixed Hamiltonians, while making explicit that the associated matter and field subsystems are gauge-relative. We compare Coulomb and multipolar descriptions under regularisation, including within the electric-dipole approximation, and identify when simplified interaction Hamiltonians remain reliable. Our analysis shows that regularisation is often benign in Coulomb-gauge treatments, but can significantly affect multipolar localisation and short-range observables when cutoffs are chosen near atomic scales. These results provide a controlled basis for judging the reliability of regularised nrQED models and indicate when beyond-dipole treatments become necessary.
We develop a regularised arbitrary-gauge formulation of non-relativistic quantum electrodynamics and use it to compare Coulomb and multipolar descriptions with a Lorentzian form factor. We analyse the effect of regularisation in perturbation theory, including alternative partitions of the Hamiltonian into free and interaction parts, and the limits of the electric dipole approximation. The regularised multipolar gauge exhibits a cut-off–dependent trade-off between the strength of individual interaction terms and the localisation of material subsystems that suppresses direct inter-atomic interactions. We discuss the implications of the framework for short-range phenomena, including Dicke criticality.
\end{abstract}

\maketitle

\section{introduction}

Non-relativistic quantum electrodynamics (nrQED) provides a rigorous framework for describing light–matter interactions at atomic and mesoscopic scales \cite{cohen-tannoudji_photons_1989,craig_molecular_1998,spohn_dynamics_2004} and its multipolar form is a central framework in light-matter physics,  with widespread applications \cite{power_coulomb_1959,lamb_fine_1952,woolley_molecular_1971,woolley_reformulation_1974,woolley_non-relativistic_1975,Javanainen_exact_2017,quesada_why_2017, stokes_implications_2021,stokes_ultrastrong_2021,stokes_gauge_2019,drezet_dual-lagrangian_2016,grinter_resonance_2016,messina_dynamical_2010,stokes_gauge_2013,stokes_noncovariant_2012,komninos_theory_2017,forbes_optical_2018,chernyak_non-linear_2015,tokman_purcell_2019,de_bernardis_breakdown_2018,stokes_extending_2012,safari_van_2008,horsley_role_2008,brooke_super-_2008,cho_single_2008,pipolo_cavity_2014}. It is immediately apparent, however, that a non-relativistic theory cannot consistently describe  interactions at relativistic energy scales. The introduction of a finite frequency cut-off of the photonic modes ensures internal consistency at the expense of an effective delocalisation of charge \cite{vukics_fundamental_2015,Ruggenthaler2023}. The consequences of such a procedure are especially important at small separations approaching the regime of atomic overlap, as occurs, for example,  for the Dicke-model's phase transition \cite{Ritsch2013RMP,grieser_depolarization_2016,stokes_uniqueness_2020,brooke_super-_2008}.
The effect is most pronounced in descriptions that would otherwise define more localised material subsystems, such as the multipolar description, wherein the finite cut-off reduces the suppression of direct inter-system interactions. A closely related topic is the consistency and validity of the electric dipole approximation (EDA), which assumes emitters are small compared to relevant resonant wavelengths.

Regularisation has been considered in specific contexts, \cite{cohen-tannoudji_photons_1989,grieser_depolarization_2016,Woolley2020,Woolley2024}, but an arbitrary-gauge formulation is lacking. Since distinct gauge choices provide physically distinct partitions of the composite system into canonical light and matter quantum subsystems, non-trivial interplay between regularisation and gauge choice is to be expected \cite{stokes_implications_2022,stokes2023gauge,Stokes2023}. In this work, we construct a regularised arbitrary-gauge formulation of nrQED, and quantify the impact of regularisation on single- and multi-atom Hamiltonians. We examine alternative partitions into free and interaction terms, and identify the conditions under which the electric dipole approximation remains reliable and compatible between Coulomb and multipolar definitions of the material subsystem. Our results clarify how regularisation modifies short-range physics and delineates the regimes in which different theoretical descriptions remain self-consistent.

\section{Arbitrary gauge regularised Hamiltonian}

\subsection{Derivation}

We consider the simplest physical system that allows us to fully understand the implications of gauge-choice, regularisation, and their interplay, for both interatomic and atomic self-interactions: two isolated, stationary hydrogen atoms. The extension to multiple arbitrary charge distributions is straightforward, but does not provide additional insight into the topics of focus here. Dynamical charges $q$ at ${\bf r}_n.~n=1,2$ are separately anchored to charges $-q$ fixed at ${\bf R}_n$ with ${\bf R}_1={\bf 0}$ and ${\bf R}_2={\bf R}$. We will assume that the atoms never comprise an H$_2$ molecule, that is, $R$ is always greater than the effective radius of a hydrogen atom when occupying a typical low-lying energy state. The point-charge densities of charge and current are
\begin{align}
\rho({\bf x}) &= \sum_{n=1,2} \rho_n({\bf x}) =\sum_{n=1,2} q\left[\delta({\bf x}-{\bf r}_{n})-\delta({\bf x-R}_{n})\right],\label{chargedens}\\
{\bf J}({\bf x}) &= \sum_{n=1,2}{\bf J}_n ({\bf x}) \nonumber \\ &=\sum_{n=1,2} {q\over 2}\left[{\dot {\bf r}_{n}}\delta({\bf x}-{\bf r}_{n})+\delta({\bf x}-{\bf r}_{n}){\dot {\bf r}_{n}}\right].\label{currentdens}
\end{align}
The convolution of an arbitrary field ${\bf F}$ with a smearing function $\varphi$ is defined by
\begin{align}\label{convIdent}
{\bf F}_\varphi({\bf x}) &:=  \int {d^3 k \over (2\pi)^3} {\bf F}_\varphi({\bf k})e^{i{\bf k}\cdot {\bf x}} = \int d^3 x' {\bf F}({\bf x}')\varphi({\bf x}-{\bf x}'), \nonumber \\
&=:[{\bf F}*\varphi]({\bf x})
\end{align}
where $ {\bf F}_\varphi({\bf k}):= \varphi({\bf k}){\bf F}({\bf k})$, and we rely entirely upon the arguments ${\bf x}$ and ${\bf k}$ to distinguish between a field and its Fourier transform. The form factor $\varphi({\bf k})$ suppresses the contributions of modes $k$ above a specified cut-off $k_\varphi$. Applied to Eqs.~(\ref{chargedens}) and (\ref{currentdens}), it determines the relative localisation of the  material fields 
\begin{align}
\rho_\varphi({\bf x})  &= \sum_{n=1,2} \rho_{\varphi n}({\bf x}) = \sum_{n=1,2} q[\varphi({\bf x}-{\bf r}_{n})-\varphi({\bf x-R}_{n})], \label{nprho}\\
{\bf J}_\varphi({\bf x}) &= \sum_{n=1,2} {\bf J}_{\varphi n}({\bf x}) \nonumber \\ &= \sum_{n=1,2} {q\over 2}\left[{\dot {\bf r}_n}\varphi({\bf x}-{\bf r}_n)+\varphi({\bf x}-{\bf r}_n){\dot {\bf r}_n}\right],\label{npcurrent}
\end{align}
with the limit $\varphi({\bf k}) \equiv 1$ yielding the point charge densities. 

We now provide an arbitrary gauge canonical quantum theory using the regularised densities in Eqs.~(\ref{nprho}) and (\ref{npcurrent}). The electric field denoted ${\bf E}$, must satisfy Gauss' law
\begin{align}
C_1 = \nabla \cdot {\bf E}-\rho_\varphi = 0
\end{align}
which is a non-dynamical constraint, while the magnetic field, ${\bf B}$, must satisfy the homogeneous Maxwell equations ${\dot {\bf B}} = -\nabla \times {\bf E}$ and $\nabla \cdot {\bf B}=0$. The latter are satisfied identically if one introduces a potential $A$ with components $(A_\mu) = (A_0,-{\bf A})$, such that ${\bf E}=-\nabla A_0 -{\dot {\bf A}}$ and ${\bf B}=\nabla \times {\bf A}$. Accordingly, the electric and magnetic fields are invariant under the gauge transformation $A_\mu \to A_\mu -\partial_\mu \chi$. 
%The remaining Maxwell-Ampere and Lorentz force laws are dynamical equations the canonical theory provides.

Any suitably well-behaved three vector field ${\bf V}$ admits a unique (Helmholtz) decomposition into transverse (divergenceless), ${\bf V}_{\rm T}$, and longitudinal (curl-free), ${\bf V}_{\rm L}$, parts, as ${\bf V}={\bf V}_{\rm T}+{\bf V}_{\rm L}$. Since $\nabla \times \nabla \chi \equiv {\bf 0}$, the transverse potential ${\bf A}_{\rm T}$ is gauge-invariant while the longitudinal potential ${\bf A}_{\rm L}={\bf A}-{\bf A}_{\rm T}$ is freely choosable.

As in Refs.~\cite{woolley_charged_1999,stokes_identification_2021}, we take the gauge-fixing constraint
\begin{align}\label{gconstraint}
C_2 = \int d^3 x' {\bf g}({\bf x}',{\bf x})\cdot {\bf A}({\bf x}')
\end{align}
where $\nabla \cdot {\bf g}({\bf x},{\bf x}')=\delta({\bf x}-{\bf x}')$. The longitudinal green's function ${\bf g}_{\rm L}$ is uniquely defined as the gradient of the green's function for the Laplacian;
\begin{align}
{\bf g}_{\rm L}({\bf x},{\bf x}')  = -\nabla {1\over 4\pi|{\bf x}-{\bf x}'|},\label{gl}
\end{align}
whereas the component ${\bf g}_{\rm T}={\bf g}-{\bf g}_{\rm L}$ is an essentially arbitrary transverse (w.r.t ${\bf x}$) field.

The constraint $C_2=0$ implies that
\begin{align}
{\bf A}({\bf x}) = {\bf A}_g({\bf x}) \equiv {\bf A}_{\rm T}({\bf x}) +\nabla \int d^3 x' {\bf g}_{\rm T}({\bf x}',{\bf x})\cdot {\bf A}_{\rm T}({\bf x}'),
\end{align}
in which the choice of gauge is determined entirely by ${\bf g}_{\rm T}$. This constraint is therefore sufficiently general to yield the most commonly chosen gauges of nonrelativistic QED as special cases \cite{stokes_implications_2022}, namely, the Coulomb gauge, ${\bf g}_{\rm T}={\bf 0}$, and the Poincar\'e (multipolar) gauge,
\begin{align}
{\bf g}_{\rm T}({\bf x},{\bf x}') &= - \int_{\bf C} d{\bf s} \cdot \delta^{\rm T}({\bf x}-{\bf s}) \nonumber \\ &= -\int_0^1 d\lambda\, {\bf x}'\cdot \delta^{\rm T}({\bf x}-\lambda{\bf x}'),\label{mpg}
\end{align}
where ${\cal C}(\lambda) = \lambda {\bf x}',~\lambda \in [0,1]$ is the straight line from ${\bf 0}$ to ${\bf x'}$. Recall here that the transverse $\delta$-function $\delta^{\rm T}$ is defined by
\begin{align}
\delta_{ij}^{\rm T}({\bf x}) = \delta_{ij}\delta({\bf x})-\delta_{ij}^{\rm L}({\bf x}) 
\end{align}
where
\begin{align}\label{dlong}
\delta_{ij}^{\rm L}({\bf x}) = -\nabla_i \nabla_j {1\over 4\pi|{\bf x}|} = \int {d^3k\over (2\pi)^3} \, {\hat k}_i{\hat k}_j e^{i{\bf k}\cdot {\bf x}}
\end{align}
defines the longitudinal $\delta$-function $\delta^{\rm L}$.

Analogous to the electromagnetic potentials, regularised auxiliary material potentials ${\bf P}_\varphi$ and ${\bf M}_\varphi$ can be defined using the {\em inhomogeneous} Maxwell equations;
\begin{align}
&\rho_\varphi = -\nabla\cdot {\bf P}_\varphi \label{P},\\
&{\bf J}_\varphi ={\dot {\bf P}}_\varphi+\nabla \times {\bf M}_\varphi.
\end{align}
Unlike ${\bf E}$ and ${\bf B}$, which must satisfy the accompanying homogeneous Maxwell equations, the material potentials ${\bf P}_\varphi$ and ${\bf M}_\varphi$ possess gauge freedom. Specifically, the physical charge and current densities $\rho_\varphi$ and ${\bf J}_\varphi$ are invariant under a transformation by pseudo-magnetic and pseudo-electric fields as
\begin{align}
&{\bf P}_\varphi\to {\bf P}_\varphi+\nabla \times {\bf U},\label{pmt}\\
&{\bf M}_\varphi \to {\bf M}_\varphi-\nabla U_0 -{\dot {\bf U}}
\end{align}
where $(U_\mu)=(U_0,-{\bf U})$ are the components of an arbitrary pseudo-four-potential. The fields ${\bf P}_\varphi$ and ${\bf M}_\varphi$ are in turn invariant under a gauge transformation $U_\mu\to U_\mu -\partial_\mu \chi$ where $\chi$ is arbitrary.

The field ${\bf M}_{\varphi\rm L}$ is completely arbitrary, such that only the transverse freedom in ${\bf P}_\varphi$ and ${\bf M}_\varphi$ is non-trivial. If we define ${\bf P}_\varphi$ by
\begin{align}\label{P2}
{\bf P}_{g\varphi}({\bf x}) &= -\int d^3 x' {\bf g}({\bf x},{\bf x}')\rho_\varphi({\bf x}') \nonumber \\ & = -\sum_{n=1,2}\int d^3 x' {\bf g}({\bf x},{\bf x}')\rho_{\varphi n}({\bf x}') = \sum_{n=1,2} {\bf P}_{g\varphi n}({\bf x}),
\end{align}
then Eq.~(\ref{pmt}) is satisfied identically with ${\bf P}_{\varphi \rm L}=\nabla \phi_\varphi$ where
\begin{align}
\phi_\varphi({\bf x}) = -{1\over \nabla^2}\rho_\varphi({\bf x})= \int d^3 x' {\rho_\varphi({\bf x}')\over 4\pi|{\bf x}-{\bf x}'|},
\end{align}
is the Coulomb-potential obtained from the expression for ${\bf g}_{\rm L}$ given in Eq.~(\ref{gl}).

The transverse polarisation ${\bf P}_{g\varphi \rm T}$ is completely arbitrary and, like ${\bf A}_g$, is fully specified by ${\bf g}_{\rm T}$. In the Coulomb gauge (${\bf g}_{\rm T}={\bf 0}$), for example, we have ${\bf P}_{\rm C\varphi} ={\bf P}_{\varphi \rm L} = -{\bf E}_{\varphi \rm L}= \nabla \phi_\varphi$.
It is easy to show via the gradient theorem or by Fourier transformation that for a $\varphi$ which vanishes at infinity ${\bf P}_{\varphi \rm L}$ can also be written as a line-integral between the two charges
\begin{align}\label{PL}
{\bf P}_{\varphi \rm L}({\bf x}) = -\int d^3 x' \int_{\cal C} d{\bf s} \cdot \delta^{\rm L}({\bf x}-\lambda{\bf x}')\rho_\varphi({\bf x'}).
\end{align}
Which, as a longitudinal field, is ${\cal C}$-path independent. If we now choose the Poincar\'e-gauge, then ${\bf g}_{\rm T}$ is given by Eq.~(\ref{mpg}) and upon using Eq.~(\ref{PL}) we obtain
\begin{align}
{\bf P}_{\rm M\varphi} ({\bf x}) = -\int d^3 x' \int_{\cal C} d{\bf s}\, \delta({\bf x}-\lambda{\bf x}')\rho_\varphi({\bf x'}).
\end{align}
In the point charge limit this is nothing but the well-known {\em multipolar} atomic polarisation field %\textcolor{red}{should the first R be there in the second term?}
\begin{align}\label{multP}
&{\bf P}_{\rm M}({\bf x}) \\&=\int_0^1 d\lambda \left[ {\bf d}_1\delta({\bf x}-\lambda{\bf r}_1)+ {\bf d}_2\delta({\bf x}-{\bf R}-\lambda[{\bf r}_2-{\bf R}])\right],
\end{align}
where ${\bf d}_1 = q{\bf r}_1$ and ${\bf d}_2= q({\bf r}_2-{\bf R})$ are the atomic dipole moments.

The canonical quantum theory is expressed in terms of canonical operators
$\{{\bf r}_n,\,{\bf p}_n, \,{\bf A}_{\rm T}, {\bf \Pi}\}$
satisfying
\begin{align}
[r_i,p_j] &=i\delta_{ij},\label{com1}\\
[A_{{\rm T}i}({\bf x}),\Pi_j({\bf x}')] &= i\delta_{ij}^{\rm T}({\bf x}-{\bf x}'),\label{com2}
\end{align}
with all other canonical commutators identically zero. Denoting by ${\bf e}_1$ and ${\bf e}_2$ the unit vectors orthogonal to ${\hat {\bf k}}$, the annihilation operator for a photon with momentum ${\bf k}$ and polarisation $\lambda$ is defined by \cite{cohen-tannoudji_photons_1989}
\begin{align}\label{phmodeeqn}
a_\lambda({\bf k})= { {\bf e}_\lambda({\bf k})\over \sqrt{2k}}\cdot [k{\bf A}_{\rm T}({\bf k}) + i {\bm \Pi}({\bf k})],
\end{align}
such that
\begin{align}
[a_\lambda({\bf k}),a^\dagger_{\lambda'}({\bf k}')] = \delta_{\lambda\lambda'}\delta({\bf k}-{\bf k}').
\end{align}
The Hilbert space and operator algebra of the theory are ${\cal H}={\cal H}_m\otimes {\cal H}_{\rm ph}$ and ${\cal A} ={\cal A}_m \otimes {\cal A}_{\rm ph}$ where ${\cal A}_s$ is the algebra of hermitian operators acting on ${\cal H}_s,~s=m,\, {\rm ph}$. The material canonical operators possess the form ${\bf r}_n\otimes  I_{\rm ph}$ and ${\bf p}_n\otimes I_{\rm ph}$, acting nontrivially only within ${\cal H}_m$, while the Maxwell canonical operators $I_m\otimes{\bf A}_{\rm T}$ and $I_m\otimes {\bf\Pi}$ act non-trivially only within ${\cal H}_{\rm ph}$.

Dirac's constrained quantisation procedure reveals how the physical observables $\{\rho_\varphi, {\bf J}_\varphi ,{\bf B},{\bf E}\}$ are related to the canonical operators $\{{\bf r},\,{\bf p}, \,{\bf A}_{\rm T}, {\bf \Pi}\}$ whose algebraic properties and whose action on the physical state space ${\cal H}$ are known. This is achieved through the construction of Dirac Brackets. Details of the calculation can be found in Ref.~\cite{stokes_identification_2021}, which considers the point charge limit, but the results of which can carried over to the present case via the replacements $\rho\to \rho_\varphi$ and ${\bf J}\to {\bf J}_\varphi$. Various observable commutation relations for this case are given in Appendix~\ref{Appendix A}. Gauss' law ($C_1=0$) implies that there necessarily exists a ``material" part, ${\bf P}_{g\varphi}$, of the electric field. Dirac's constrained quantisation procedure reveals that the remaining part is nothing but (minus) the photonic momentum $-{\bf \Pi}$ in Eq.~(\ref{com2});
\begin{align}\label{Efield}
{\bf E} = -{\bf \Pi}-{\bf P}_{g\varphi}.
\end{align}
Since $C_1=0$ implies ${\bf E}_{\rm L} = -{\bf P}_{\varphi \rm L}$, the field ${\bf \Pi} = -{\bf E}_{\rm T}-{\bf P}_{g\varphi\rm T}$ is transverse and possesses a different physical meaning for each different choice of ${\bf g}_{\rm T}$. Similarly, the material canonical momenta ${\bf p}_n$ are related to the corresponding manifestly gauge-invariant mechanical momenta $m{\dot {\bf r}}_n$ via minimal coupling;
\begin{align}\label{minimal coupling ansatz}
m{\dot {\bf r}}_n = {\bf p}_n- q{\bf A}_{g\varphi}({\bf r}_n),
\end{align}
such that ${\bf p}_n$ is physically distinct for each different choice of ${\bf g}_{\rm T}$. The final Hamiltonian on the physical state space is the total energy,
\begin{align}\label{en}
H_g =&\sum_{n=1,2} {1\over 2}m{\dot {\bf r}}_n^2 +{1\over 2}\int d^3 x \left[{\bf E}({\bf x})^2+{\bf B}({\bf x})^2\right],
\end{align}
which in terms of the canonical momenta reads
\begin{align}\label{ham}
H_g =&{1\over 2m}\sum_{n=1,2}\left[{\bf p}_n-q{\bf A}_{g\varphi}({\bf r}_n)\right]^2 \nonumber \\&+{1\over 2}\int d^3 x \left(\left[{\bf \Pi}({\bf x})+{\bf P}_{g\varphi}({\bf x})\right]^2+{\bf B}({\bf x})^2\right),
\end{align}
where ${\bf B}=\nabla \times {\bf A}_g =\nabla \times {\bf A}_{\rm T}$.

The equation of motion of an arbitrary operator $O\{{\bf r}_n,\,{\bf p}_n, \,{\bf A}_{\rm T},{\bf \Pi}\}$ is found using the Heisenberg equation ${\dot O}=-i[O,H_g]$ and the canonical commutation relations in Eqs.~(\ref{com1}) and (\ref{com2}). In particular, we verify in the Appendix that the Hamiltonian $H_g$ in Eq.~(\ref{ham}) yields the required Maxwell-Ampere law, ${\dot {\bf E}}=\nabla\times {\bf B}-{\bf J}_\varphi$, and regularised Newton-Lorentz force law, $m{\ddot {\bf r}}_n = q[{\bf E}_\varphi({\bf r}_n)+\{{\dot {\bf r}}_n \times {\bf B}_\varphi ({\bf r})-{\bf B}_\varphi({\bf r}_n)\times {\dot {\bf r}}_n\}/2]$, for any choice of gauge ${\bf g}_{\rm T}$, and any choice of form factor $\varphi$.

The Hamiltonians defined relative to different gauges ${\bf g}_{\rm T}$ and ${\bf g}_{\rm T}'$ are related by the unitary gauge-fixing transformation
\begin{align} \label{gaugefix}
U_{gg'}=\exp\left(i\int d^3x \left[{\bf P}_{g\varphi}({\bf x})-{\bf P}_{g'\varphi}({\bf x})\right]\cdot {\bf A}_{\rm T}({\bf x})\right)
\end{align}
as
\begin{align}
U_{gg'}H_gU_{gg'}^\dagger = H_{g'}.
\end{align}
Since $U_{gg'}$ is unitary, the physical content of the theory is preserved when moving between distinct fixed gauges. This is nothing but canonical nonrelativistic QED's expression of {\em gauge-invariance}. The transformation $U_{gg'}$ is, however, ``non-local" with respect to the Hilbert space tensor-product structure ${\cal H}_m\otimes {\cal H}_{\rm ph}$, that is, $U_{gg'}$ does not posses the form $U_m\otimes U_{\rm ph}$. The canonical quantum subsystems termed ``matter" and ``light" (``photons") are gauge {\em relative} \cite{stokes_implications_2022}, meaning they are physically distinct in each gauge.

In practice, the Hamiltonian is partitioned into an unperturbed part (free part) and a perturbation (interaction part) as
\begin{align}\label{freepartition}
H_g=& h+ V \equiv H_m \otimes I_{\rm ph} +I_m \otimes H_{\rm ph}+V_g.
\end{align}
The strategy invariably employed to understand quantum light-matter physics is to take the basis of eigenvectors of $h$ as a working starting point and then systematically determine the effects of the perturbation $V$.

The standard definition of the photonic Hamiltonian is
\begin{align}
H_{\rm ph} =& {1\over 2}\int d^3 x \left[{\bf \Pi}({\bf x})^2+{\bf B}({\bf x})^2\right] \nonumber \\ =& \int d^3 k \, \omega\left(a_\lambda^\dagger({\bf k})a_\lambda({\bf k})+{1\over 2}\right),
\end{align}
unless certain approximations are performed, most notably the electric dipole (long-wavelength) approximation (EDA), there is essentially no other available definition of $H_{\rm ph}$ (cf. Eqns.~(\ref{phmodeeqn}, \ref{Efield})). In contrast, the material Hamiltonian $H_m$ is not unique, even when considering only a single atom. We determine the implications of this in detail in \ref{single atom material ham section}.

\begin{figure}[h]
\centering
\resizebox{\columnwidth}{!}{
\begin{tikzpicture}[
    x=3cm, y=1cm,
    line/.style = {line width=0.9pt},
    tick/.style = {line width=0.9pt},
    lbl/.style  = {inner sep=3pt, font=\Large},
]
\draw[line] (-0.5,0) -- (3.5,0);
\draw[line] (3.5,0) -- ++(-0.20, 0.12); 
\draw[line] (-0.5,0) -- ++(0.20,-0.12);
\node[lbl, below=6pt] at (-0.5,0) {$[{\rm Length}]$};
\node[lbl, above=6pt] at (-0.5,0) {$0$};
\node[lbl, above=6pt] at (3.5,0) {$[{\rm Energy}]$};
\node[lbl, below=6pt] at (3.5,0) {$0$};
\foreach \x/\above/\below in {
  0.5/{$\alpha^{2}_{\rm fs}m_{e}\sim \omega_0$}/{$a_0/\alpha_{\rm fs}$},
  1.50/{$\alpha_{\rm fs}m_{e}\sim k_{\ell}$}/{${a_0}$},
  2.50/{$m_{e}\sim k_{m}$}/{$\alpha_{\rm fs}a_0\sim \lambda_{c}$}
}{
  \draw[tick] (\x,-0.22) -- (\x,0.22);
  \node[lbl, above=6pt] at (\x, 0.22) {\above};
  \node[lbl, below=6pt] at (\x,-0.22) {\below};
}
\end{tikzpicture}
}
\caption{Energy/length scale diagram. The horizontal axis links characteristic energy (top) and length (bottom) scales in nrQED in units such that $c=1$. The Bohr radius is denoted $a_0$, the Compton wavelength $\lambda_c$, and the electronic rest energy $m_e$. The energy ($\omega_{0}\!\sim\!\alpha_{\mathrm{fs}}^{2}m_{e}$) is a characteristic atomic ionisation transition (Lyman-$\alpha$).}
\label{fig:scale hierarchy}
\end{figure}
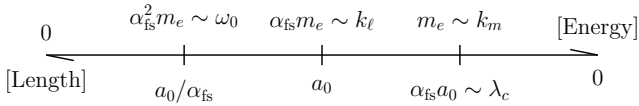

To understand the different regularisations we will consider, it is helpful to note three distinct energy scales, which we refer to as relativistic, atomic, and ionisation Fig.~(\ref{fig:scale hierarchy}). Each is obtained from the preceding one through reduction by a factor of $\alpha_{\rm fs}$ (fine-structure). The corresponding length scales therefore increase by a factor of $\alpha_{\rm fs}^{-1}$.

\subsection{Regularised dyadics}
It is helpful for what follows to provide expressions for the regularised $\delta$-function and its transverse and longitudinal dyadic variants \cite{cohen-tannoudji_photons_1989}. The Lorentzian
\begin{align}
F(k)={k_F^2\over k^2+k_F^2},
\end{align}
is a standard choice of regulator, which gives the following smeared $\delta$-functions
\begin{align}
\delta_F({\bf x}) &= \int {d^3k\over (2\pi)^3}F(k)e^{i{\bf k}\cdot{\bf x}}= {k_F^2\over 4\pi x}e^{-k_Fx},\\
\delta_{F^2}({\bf x}) &= {k_F^3\over 8\pi}e^{-k_Fx}.
\end{align}
The corresponding longitudinal and transverse dyadics are defined by
\begin{align}
\delta_F^{\rm L}({\bf x}) &= \int {d^3k\over (2\pi)^3}{\bf {\hat k}}\otimes {\bf {\hat k}}F(k)e^{i{\bf k}\cdot{\bf x}}\nonumber \\ &=-\nabla \otimes \nabla\int {d^3k\over (2\pi)^3}{F(k)\over k^2}e^{i{\bf k}\cdot{\bf x}},\\
\delta_F^{\rm T}({\bf x}) &= I_3\delta_F({\bf x})-\delta_F^{\rm L}({\bf x})
\end{align}
where $I_3$ is the three-dimensional identity.
Explicitly, we obtain the following useful expressions
\begin{align}
\delta_F^{\rm L}({\bf x}) &=-I {\tau_F'(x) \over x} - {\hat {\bf x}}\otimes {\hat {\bf x}}\left( \tau_F''(x)-{\tau_F'(x)\over x}\right),\\
\delta_{F^2}^{\rm L}({\bf x}) &=-I {\tau_{F^2}'(x) \over x} - {\hat {\bf x}}\otimes {\hat {\bf x}}\left( \tau_{F^2}''(x)-{\tau_{F^2}'(x)\over x}\right),
\end{align}
where
\begin{align}
\tau_F(x) &:=\int {d^3k\over (2\pi)^3}{F(k)\over k^2}e^{i{\bf k}\cdot{\bf x}}={1-e^{-k_F x}\over 4\pi x}, \\
\tau_{F^2}(x) &:=\int {d^3k\over (2\pi)^3}{F(k)^2\over k^2}e^{i{\bf k}\cdot{\bf x}}=\frac{2-e^{-k_F x} (2+k_F x)}{8 \pi  x}
.%,\\
%\tau'(x) &= -{1\over 4\pi x^2}\left[1-e^{-k_F x}(1+k_F x)\right],\\
%\tau''(x) &= {1\over 2\pi x^3}\left[1-e^{-k_F x}\left(1+k_F x+{1\over 2}k_F^2 x^2\right)\right],\\
%\tau'''(x) &= -\frac{6}{4 \pi  x^4}\bigg[1\nonumber \\ &~~~ -e^{-k_F x} \left\{1+k_F x+k_F^2 x^2 \left(\frac{1}{2}+\frac{1}{6}k_F x\right)\right\}\bigg].
\end{align}

\subsection{Single-atom material Hamiltonian}\label{single atom material ham section}
The electric energy, as appears in Eq.~(\ref{ham}): 
\begin{align}
{1\over 2}\int d^3 x {\bf E}({\bf x})^2 = {1\over 2}\int d^3 x \left[{\bf \Pi}({\bf x})+{\bf P}_{g\varphi}({\bf x})\right]^2,
\end{align}
possesses the purely material part
\begin{align}\label{Ug}
U_{g\varphi} = {1\over 2}\int d^3 x\, {\bf P}_{g\varphi}({\bf x})^2 = \sum_{n=1,2} U_{g\varphi n} + U_{g\varphi 12}.
\end{align}
Where the individual atomic energies are
\begin{align}
U_{g\varphi n} = {1\over 2}\int d^3 x\, {\bf P}_{g\varphi n}({\bf x})^2,
\end{align}
and the direct interatomic interaction is
\begin{align}\label{Ug12}
U_{g\varphi 12} = \int d^3 x\, {\bf P}_{g\varphi 1}({\bf x}) \cdot {\bf P}_{g\varphi 2}({\bf x}).
\end{align}
%In the Coulomb gauge the inter-atomic Coulomb interaction
%\begin{align}
%U_{\rm C\varphi 12} = \int d^3 x\, {\bf E}_{\varphi {\rm L} 1}({\bf x}) \cdot {\bf E}_{\varphi {\rm L}2}({\bf x}).
%\end{align}
%is typically included within the perturbation $V$, though it should be noted that for sufficiently small separations $R$, when dealing with atoms or molecules possessing large dipole moments, it may become necessary to include $U_{\rm C,\varphi 12}$ within the unperturbed Hamiltonian to obtain a weak perturbation \cite{}. In the multipolar gauge the polarisation field ${\bf P}_{g\varphi n}$ is much more localised in the vicinity of ${\bf R}_n$ and in the point charge limit, $\varphi({\bf k})\to 1$, it vanishes for all ${\bf x}$ not on the straight line from ${\bf R}_n$ to ${\bf r}_n$. As a result, for separate atoms, the resulting direct interaction $U_{g\varphi 12}$ vanishes identically. 
The latter will be considered in detail in Sec.~\ref{2 atom direct interaction section}. We here restrict our attention to a single atom at ${\bf 0}$ and in this section omit the atomic label ``1". Assuming that the total atomic electric potential energy is to be included within the unperturbed Hamiltonian yields the definition
\begin{align}
H_m  =  {{\bf p}^2\over 2m} + U_{g\varphi}({\bf r}).\label{hmg}
\end{align}   
In the Coulomb gauge (${\bf g}_{\rm T} \equiv {\bf 0}$) the potential energy is the atomic Coulomb energy
\begin{align}
&U_{C\varphi}({\bf r}) = {1\over 2}\int d^3 x\, {\bf E}_{\varphi {\rm L}}({\bf x})^2, 
\end{align}
which in the point charge limit $\varphi({\bf k})\to 1$ reads
\begin{align}\label{Coulomb Potential}
U_{\rm C}({\bf r})= U_{\rm C, self} - {q^2 \over 4\pi r}.
\end{align}
The second term is the familiar Coulomb binding energy between the opposite atomic charges while $U_{\rm C, self}$ denotes the Coulomb self-energies of the individual charges, which are infinite and typically ignored. Indeed, regularisation $\varphi({\bf k})\neq 1$ that suppresses high $k$ contributions renders the self-energies finite, constant and subsequently ignorable. The eigenvalue problem 
\begin{align}
\left({{\bf p}^2\over 2m} + U_{\rm C}({\bf r})\right) \psi_{nlm}({\bf r}) = \epsilon_n \psi_{nlm}({\bf r})\label{hm1}
\end{align}
is familiar to every student of quantum physics, and the solution can be found in any introductory quantum mechanics text. More general and much less studied, however, is the arbitrary atomic potential energy $U_{g\varphi}({\bf r})$, which can be partitioned as
\begin{align}
U_{g\varphi} = U_{\rm C \varphi}+\delta U_{g\varphi}
\end{align}
where
\begin{align}
\delta U_{g\varphi} = {1\over 2} \int d^3 x {\bf P}_{g\varphi \rm T}({\bf x})^2.
\end{align}
Although gauges other than the Coulomb gauge (most notably the multipolar gauge) are routinely employed in applications, the definition of material potential universally employed is $U_{\rm C}({\bf r})$. 
%In order to then obtain the same $S$-matrix in some other gauge ${\bf g}_{\rm T}\neq {\bf 0}$, the potential difference $U_{g',\rm trans}({\bf r})$ must be included within the interaction $V$. More generally, if $H_m$ is defined to possess potential energy $U_g({\bf r})$, then in any other gauge ${\bf g}'_{\rm T}$ this definition requires the inclusion of the potential difference $U_{g',\rm  trans}({\bf r})-U_{g,\rm trans}({\bf r})$ within the interaction $V$.
Indeed, a difficulty is encountered in the multipolar gauge description of point charges because $U_g({\bf r})=U_{\rm M}({\bf r})$ is in this case infinite. If one instead defines $H_m$ to possess the potential energy $U_{\rm C}({\bf r})$ then the infinity is transferred into the perturbation $V_{\rm M}$ via the difference $U_{\rm M}({\bf r})-U_{\rm C}({\bf r}) = \delta U_{\rm M}({\bf r})$. In practice, {\em renormalisation} is the accepted solution to this equivalent problem. The net result is that standard low-order perturbation theoretic $S$-matrix elements calculated using the multipolar and Coulomb gauges are finite and identical (gauge non-relative). Crucially, however, it must be recognised that this $S$-matrix is specific to having employed the definition $U_{\rm C}({\bf r})$ as the material potential energy within $H_m$.

Since the apparent necessity of renormalisation is certainly not exclusive to the multipolar, gauge nor even to a non-relativistic description of matter, the divergence of $U_{\rm M}({\bf r})$ should not be interpreted as a specific shortcoming of the standard multipolar theory. Rather, it can be viewed as a consequence of retaining arbitrarily high frequencies. Indeed, in this limit the Coulomb potential energy $U_{\rm C}({\bf r})$ does not in principle fare any better, because it too includes a divergent term, $U_{\rm C, self}$, associated with mass renormalisation.

At the length scale of an individual atom the instantaneous binding potential $U_{C}({\bf r})$ should offer an excellent approximation of the dominant near-field component of the inter-charge retarded electromagnetic interaction. Using this potential to define $H_m$ is therefore natural and allows one to study atomic physics at low energies using Coulomb-bound unperturbed states that are then weakly perturbed by $V_g$, giving ``radiative" corrections.

Within a properly regularised theory however, the arbitrary energy $U_{g\varphi}$ is generally finite. If, moreover, $\delta U_{g\varphi}$ is a weak perturbation of $U_{C\varphi}$, of the same order as standard radiative corrections, then using $U_{g\varphi}$ to define $H_m$ cannot be ruled out {\em ab initio}. The conditions under which $\delta U_{g\varphi}$ is indeed ``weak" were determined by Vukic {\em et al.} in Ref.~\cite{vukics_fundamental_2015} for a particular definition of the multipolar gauge in the long wavelength limit (regularised dipole gauge). More generally, we determine without approximation how regularisation and the additional contribution $\delta U_g$ alter the bare hydrogen atom eigenvalue problem for noteworthy choices of ${\bf g}_{\rm T}$.

\subsubsection{Coulomb gauge}

The natural definition for comparison is the well-known material Hamiltonian in Eq.~(\ref{hm1}) that results from the choices $\varphi\doteq 1$ and ${\bf g}_{\rm T}={\bf 0}$. In units such that $q^2/(4\pi\epsilon_0)=1$ the standard Coulomb potential therein is $U_{\rm C}({\bf r}) = -1/r$ (ignoring the infinite self-term). More generally, a prototypical (spherically symmetric) choice of form factor is the Lorentzian \cite{cohen-tannoudji_photons_1989,vukics_fundamental_2015}
\begin{align}
&\varphi({\bf k}) = {k_\varphi^2\over k^2+k_\varphi^2},%,\qquad &({\rm Lorentzian})
\label{reg1}
%&\varphi_2({\bf k}) = e^{-{k \over k_\varphi}}, &({\rm Exponential})\label{reg2}\\
%&\varphi_3({\bf k}) = e^{-{k^2\over k_\varphi^2}}. &({\rm Gaussian})\label{reg3}
\end{align}
which suppresses contributions of wave-vectors ${\bf k}$ with $k>k_\varphi$, and tends to unity when $k_\varphi\to \infty$.
%Using (\ref{gl}) and the convolution identity (\ref{convIdent}), we can immediately note that regularising the longitudinal electric field by allowing the charges to have finite extent is identical to regularising the longitudinal Green's function. However, for the transverse case this is no longer true. 
The resulting regularised Coulomb potential energy with $q^2/(4\pi\epsilon_0)=1$ is found to be
\begin{align}
&U_{\rm C\varphi}({\bf r}) = \frac{1}{r}\left(e^{-k_\varphi r}-1\right)+\frac{k_\varphi}{2} \left(e^{-k_\varphi r}+1\right).\label{uc1}
%&U_{\rm Coul,2}({\bf k}) = -\frac{2}{\pi r} \tan ^{-1}\left(\frac{k_\varphi r}{2}\right),\label{uc2}\\
%&U_{\rm Coul,3}({\bf k}) = -\frac{1}{r}\text{erf}\left(\frac{k_\varphi r}{2 \sqrt{2}}\right),\label{uc3}
\end{align}
For $k_\varphi r\gg1$ this energy behaves as $-1/r$, while for $k_\varphi r\sim 1$ a departure from this behaviour is observed as the form factor removes the singularity at the origin. Thus, for $k_\varphi \sim a_0^{-1}$ (inverse Bohr radius) one expects significant deviations in the spectrum of $H_m$ from the standard spectrum obtained from Eq.~(\ref{hm1}), namely, $\epsilon_n=-1/(2n^2)$. On the other hand, for $k_\varphi$ of the order of the inverse Compton wavelength, which is a natural nonrelativistic cutoff (Fig.~\ref{fig:scale hierarchy}), the standard spectrum is essentially recovered.

\subsubsection{Multipolar gauges}\label{MultipolarGaugeSection}
%\begin{figure}[h]\label{efunc1}
%\centering
%\includegraphics[width=\linewidth]{figures/efunc1.jpg}
%\caption{Radial eigenfunctions against `$|\mathbf{r}|$' with $H_m$ possessing the potential energy $U_{M_{\ell}}$ for $n=1,2,3$ (black, blue, green) are shown dashed, the solid lines correspond to the analytic eigenfunctions when $H_m$ possesses solely the $-1/r$ potential.}
%\label{fig:efunc}
%\end{figure}
The standard multipolar gauge choice of ${\bf g}_{\rm T}$ in Eq.~(\ref{mpg}) yields,
in the point charge limit $k_\varphi\to \infty$ (with $q^2/(4\pi\epsilon_0)=1$),
\begin{align}
\delta U_{\rm M}({\bf r}) = {2\pi} \int_0^1 d\lambda d\lambda' {\bf r}\cdot \delta^{\rm T}([\lambda-\lambda']{\bf r})\cdot {\bf r}.
\end{align} 
The integrand is singular at contact points $\lambda=\lambda'$.
%We define smeared dyadics
%\begin{align}\label{multgT2}
%\delta_{\ell ij}^{\rm L}({\bf x}) &= \int {d^3k\over (2\pi)^3} \ell({\bf k}){\hat k}_i{\hat k}_je^{i{\bf k}\cdot {\bf x}}\\
%\delta_{\ell ij}^{\rm T}({\bf x}) &= \int {d^3k\over (2\pi)^3} \ell({\bf k})(\delta_{ij}-{\hat k}_i{\hat k}_j)e^{i{\bf k}\cdot {\bf x}} \\
%\delta_{ij}\delta_\ell({\bf x}) &= \delta_{\ell ij}^{\rm L}({\bf x})+\delta_{\ell ij}^{\rm T}({\bf x}),
%\end{align}
%which generalise the corresponding ordinary $\delta$-function dyadics via the inclusion of the form factor $\ell$.
A finite term, $\delta U_{\rm M_\ell}$, is obtained in the point charge limit for the regularised multipolar-gauge choice
\begin{align}\label{multgT}
{\bf g}_{\ell\rm T}({\bf x},{\bf x}') =-\int_0^1 d\lambda\, {\bf x}'\cdot \delta_\ell^{\rm T}({\bf x}-\lambda{\bf x}')
\end{align}
For concreteness, we again consider the Lorentzian form factor $\ell({\bf k}) = k_\ell^2/(k^2+k_\ell^2)$ with cut-off $k_\ell$, which gives the total potential energy
\begin{align}
\delta U_{{\rm M}_\ell}({\bf r}) &= k_\ell\left(\frac{k_\ell r}{2} - 1\right)-\frac{1}{r}\left(e^{-k_\ell r}-1\right),\label{mul1}\\
U_{{\rm M}_\ell}({\bf r}) &= \delta U_{{\rm M}_\ell}({\bf r})-{1 \over r}.\label{mul2}
\end{align}
The bare Coulomb term $-1/r$ is exactly cancelled in $U_{{\rm M}_\ell}$ by an equal and opposite term coming from $\delta U_{{\rm M}_\ell}$, in agreement with Woolley's analysis in Ref.~\cite{woolley_power-zienau-woolley_2020}. The linear term $k_\ell^2 r/2$ dominates for $k_\ell r\gg 1$, whereas the Coulomb-like term $-e^{-k_\ell r}/r$ in Eq.~(\ref{mul2}) tends to the bare counterpart $-1/r$ for sufficiently small $k_\ell r$. The quadratic electric dipole approximation $\delta U_{{\rm M}_\ell} \approx k_\ell^3 r^2/6$ found by Vukic {\em et al.} in Ref.~\cite{vukics_fundamental_2015} is recovered from Eq.~(\ref{mul2}) by expanding $e^{-k_\ell r}$ to cubic order \footnote{To be precise, changing units by reinserting a factor of $q^2/(4\pi\epsilon_0)$ yields the result given in Ref.~\cite{vukics_fundamental_2015}, namely, $\delta U_{{\rm M}_\ell}({\bf r})=k_\ell ^3 d^2/(24\pi\epsilon_0)$.}.

Following Ref.~\cite{vukics_fundamental_2015}, we refer to a requirement that $\delta U_{g\varphi}$ is a weak perturbation of the bare Coulomb energy $U_{\rm C}({\bf r})=-1/r$, as {\em condition I}. In the EDA, $U_{{\rm M}_\ell} \approx k_\ell^3 r^2/6$, it is satisfied provided $(k_\ell a_0)^3 \ll 1$ where $a_0$ is the Bohr radius, which is consistent with the expansion that yields the EDA result. The cubic power implies that even for $k_\ell$ as large as $a_0^{-1}/2$ one still obtains $(k_\ell a_0)^3\sim 0.1$ \cite{vukics_fundamental_2015}. Since condition I requires that $k_\ell$ is not larger than the inverse atomic size, $k_\ell \lesssim a_0^{-1}$, it cannot be satisfied for the natural non-relativistic cut-off that we considered in the Coulomb gauge, namely, $\lambda_c^{-1} = (\alpha_{\rm fs} a_0)^{-1} \gg a_0^{-1}$ where $\alpha_{\rm fs}$ is the fine structure constant. For such a cut-off, $k_\ell r \gg 1$, and the Coulomb-like term in Eq.~(\ref{mul2}) vanishes, so $U_{{\rm M}_\ell}$ bears essentially no resemblance to the bare Coulomb potential $-1/r$. 
%The low energy spectrum of $H_m$ with potential energy $U_{{\rm M}_\ell}$, and the radial parts of the corresponding eigenfunctions are shown in Fig.~\ref{fig:efunc}.

\begin{figure}[t]
\centering
\includegraphics[width=\linewidth]{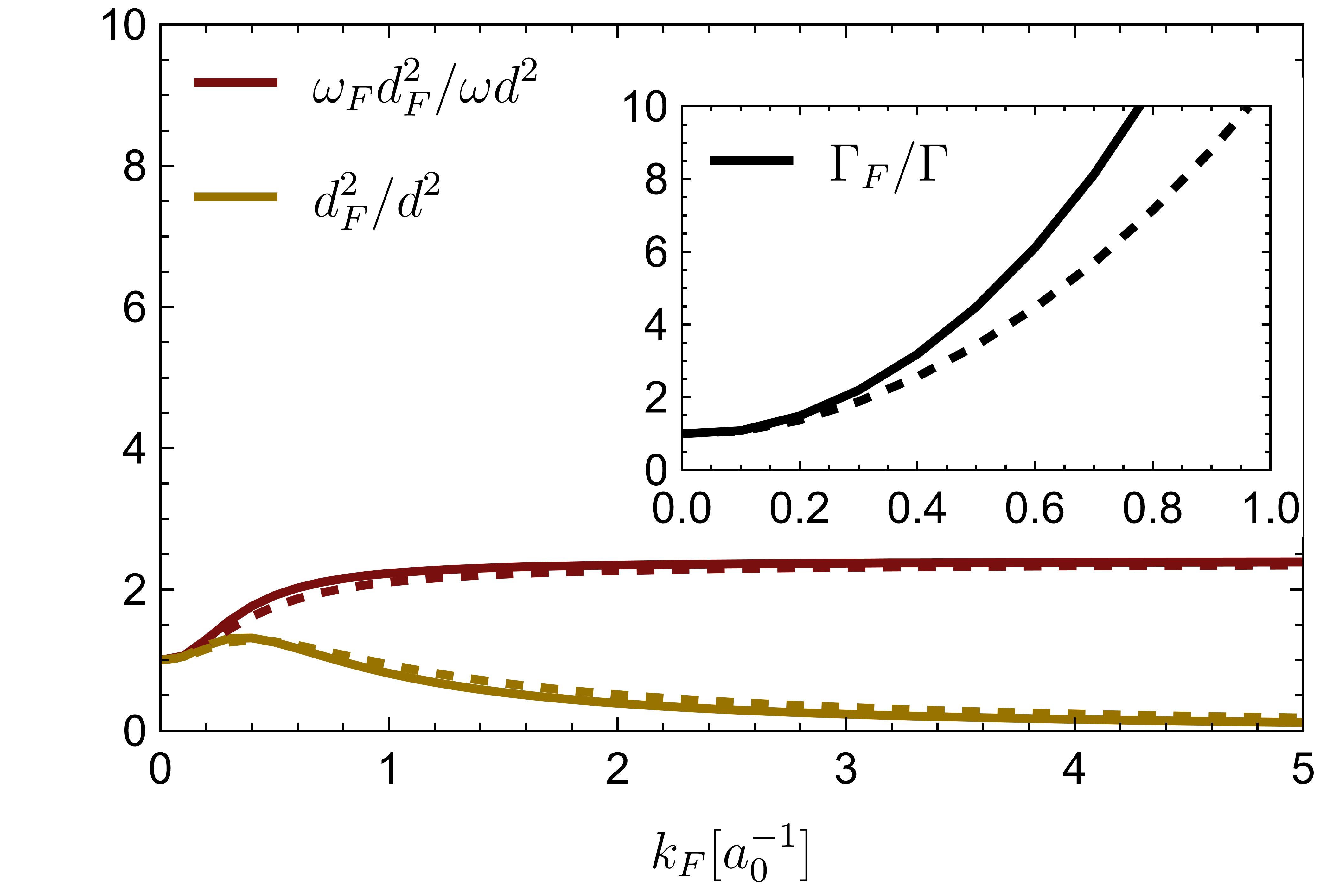}
\caption{Main: $d^{2}_{F}/d^{2}$ (dark yellow), $\omega_{F} d^{2}_{F}/\omega d^{2}$ (dark red), against $k_{F}$; $d$ and $\omega$ denote the $2p\to 1s$ transition dipole moment and transition frequency respectively, while $d_F$ and $\omega_F$ denote their $k_F$-dependent counterparts obtained when $\delta U_{M_F}$ is included within $H_m$. Dashed lines are obtained from the dipole approximation of $\delta \tilde{U}_{M\varphi}$ and solid lines from the unapproximated $\delta U_{M_{F}}$. A maximum value of $d^{2}_{0.36}/d^{2}\approx 1.32$ is followed by a linear decline as confinement increases with $k_F$. This is compensated by a corresponding growth in $\omega_{F}$. Inset:  
Eq.~(\ref{gammaratio}) as a function of $k_{F}$, shown in units of the bare rate. Significant deviation of $\Gamma_F/\Gamma$ from unity is seen even for relatively modest cut-off values, $k_{F}\sim1/2 [a_0^{-1}]$.}
\label{figgammaratio}
\end{figure}

To determine the effect of such a redefinition of $H_m$ on physical predictions, we consider the dipolar spontaneous emission rate of the $2p\to 1s$ transition. We let $\omega$ and $d$ denote the $2p\to 1s$ transition energy and transition dipole moment when $H_m$ possesses the usual bare Coulomb potential energy $U_C({\bf r})=-1/r$, giving the rate $\Gamma = \omega^3 d^2/(3\pi) = (2/3)^8\alpha_{\rm fs}^5 m$. We let $\omega_\ell$, $d_\ell$, and $\Gamma_\ell$ denote the counterparts when $H_m$ instead possesses the potential energy $U_{{\rm M}_\ell}$. The ratio
\begin{align}\label{gammaratio}
{\Gamma_\ell \over \Gamma} = {\omega_\ell^3 d_\ell^2 \over \omega^3 d^2}
\end{align}
is shown in Fig.~\ref{figgammaratio}. Another obvious second-order prediction to consider is the Lamb shift, but we postpone doing so until Sec.~\ref{full}.

The regularised multipolar gauge can straightforwardly be generalised via the choice
\begin{align} \label{convG}
{\bf g}_{\ell \varphi \rm T}({\bf x},{\bf k}) ={{\bf g}_{\ell \rm T}({\bf x},{\bf k})\over \varphi({\bf k})}.
\end{align}
Substitution into Eq.~(\ref{P2}) shows that the resulting polarisation is identical to that obtained from Eq.~(\ref{multgT}) and the density of point charges, viz.,
\begin{align}
{\bf P}_{\ell\varphi\rm T}({\bf x}) &= -\int d^3 k\, {{\bf g}_{\ell \rm T}({\bf x},{\bf k})^*\over \varphi({\bf k})} \rho_\varphi({\bf k}) \nonumber \\ &= -\int d^3 k \,{\bf g}_{\ell \rm T}({\bf x},{\bf k})^* \rho({\bf k}) = {\bf P}_{\ell\rm T}({\bf x}).
\end{align}
Thus, the Coulomb energy $U_{C\varphi}$ and the correction $\delta U_{g\varphi}$ are independently regularised by Lorentzian functions $\varphi$ and $\ell$ with respective cut-offs $k_\varphi$ and $k_\ell$. The total potential energy is the sum of Eqs.~(\ref{uc1}) and (\ref{mul1}), which ignoring constant terms reads
\begin{align}\label{Upzw}
U_{{\rm M}_{\varphi\ell}}({\bf r}) =& \frac{1}{r}(e^{-k_\varphi r}-e^{-k_\ell r})+{1\over 2}\left(k_\varphi e^{-k_\varphi r}+ k_\ell^2 r\right).
\end{align}
For $k_\varphi = k_\ell$ the highly non-Coulomb-like (confining) potential, $k_\varphi \left(e^{-k_\varphi r}+ k_\varphi r\right)/2$, is obtained. If, in contrast, the two cut-offs are separated by at least a factor of $\alpha_{\rm fs}$ as $k_\varphi \sim \lambda_c^{-1}$ and $k_\ell \lesssim a_0^{-1}$, then $U_{\rm C\varphi}$ is essentially the bare Coulomb potential $U_{\rm C}$  (strictly recovered when $k_\varphi \to \infty$), while $\delta U_{{\rm M}_\ell}$ is a weak perturbation.   

\subsubsection*{Discussion}

We have analysed the difference $\delta U_{g\varphi}$ between the total material electric energy and the Coulomb energy, for different variants of the multipolar gauge, consistent with but extending analyses elsewhere \cite{vukics_fundamental_2015,woolley_power-zienau-woolley_2020}. When included in the unperturbed Hamiltonian, this difference must be a weak perturbation of the standard $H_m$ to avoid a drastic departure from the well-known and ubiquitously used unperturbed material basis resulting from the Coulomb energy (condition I).  A cut-off no larger than half the inverse Bohr radius is required for condition I to hold. 
The question of whether to include $\delta U_{g\varphi}$ within $H_m$ is of particular importance for understanding criticality in many-body cavity QED systems \cite{vukics_adequacy_2012,grieser_depolarization_2016,stokes_uniqueness_2020,brooke_super-_2008,de_bernardis_cavity_2018}. A modest cut-off $k_\ell \sim a_0^{-1}$ should be expected to result in a delocalisation of the multipolar polarisation significantly greater (though still exponential) than occurs for the natural non-relativistic cut-off $\lambda_c^{-1} \gg a_0^{-1}$. This relative delocalisation may have a significant effect at the high densities required for a quantum phase transition. For example, in Ref.~\cite{grieser_depolarization_2016} it was found within the EDA that the weak $\delta U_{g\varphi}$ results in a ``depolarisation shift", increasing the critical density for ground state superradiance, which would then begin to border the density at which dipolar overlap and crystalisation occurs.

The relative localisation of the material polarisation and its consequences for inter-atomic interactions are discussed in Sec.~\ref{2 atom direct interaction section}. We note first that it is essential to recognise that the partition $H_g=h+V_g$ in Eq.~(\ref{freepartition}) is nothing more than a calculational tool, chosen for utility and convenience. While the operators $H_m, \, H_{\rm ph}$, and $V_g$ are always expressible in terms of the primitive manifestly gauge-invariant and local observables $\{\rho_\varphi, {\bf J}_\varphi ,{\bf B},{\bf E}\}$, they are not required individually to possess any special physical relevance. Indeed, the partition in Eq.~(\ref{freepartition}) is necessarily distinct from the partition in Eq.~(\ref{en}) into true kinetic and electromagnetic energies, which to work with directly would require a quite different approach \cite{stokes_nonconjugate_2022}.

There is nothing to prohibit, if it is useful, the inclusion within $V_g$ of purely material or photonic operators with the form $O_m \otimes I_{\rm ph}$ and $I_m\otimes O_{\rm ph}$. Indeed, we have seen that even if $\delta U_{g\varphi}$ is weak, its modification of $H_m$ can have significant impact (Fig.~\ref{figgammaratio}). To avoid such an outcome, one must include $\delta U_{g\varphi}$ within $V_g$. If, in doing so, $\delta U_{g\varphi}$ is weak, then the expected and usually required weakness of $V_g$ is not violated. But weakness of $\delta U_{g\varphi}$ still requires a modest cut-off, $k_\ell \sim a_0^{-1}$, compromising the spatial localisation of the multipolar polarisation, which is a key practical advantage of the framework \cite{cohen-tannoudji_photons_1989,stokes_implications_2022,craig_molecular_1998}. In what follows we focus on the most commonly used forms of nrQED. We show in particular that it is possible to define a practical dipole gauge theory including both the standard definition of $H_m$ and the natural non-relativistic cut-off $\lambda_c^{-1} \gg a_0^{-1}$, but which nevertheless possesses a {\em weak} perturbation $V_g$.   

\section{Single atom Coulomb and Multipolar gauge Hamiltonians}\label{full}

\subsection{Electric dipole approximation}

\subsubsection{Coulomb gauge}\label{couleda}

We denote by $\lambda_0$ a relevant resonant wavelength and then suppose that $k_\varphi \sim \lambda_c^{-1}$, so the energy scale of interest spans $\lambda_0^{-1}$ up to $\lambda_c^{-1}$ (Fig.~\ref{fig:scale hierarchy}). If $\lambda_0$ satisfies $\lambda_0 \gg a_0 \sim r \gg \lambda_c$, the variation in ${\bf A}_{\rm T\varphi}({\bf r})$ with ${\bf r}$ is ignorable. We then have ${\bf A}_{\rm T\varphi}({\bf r})\approx {\bf A}_{\rm T\varphi}({\bf 0})$, which is the EDA. The single atom, Coulomb gauge (${\bf g}_{\rm T}={\bf 0}$), regularised Hamiltonian then becomes
\begin{align}
H_{\rm C}^\varphi = {1\over 2m}\left[{\bf p}-q{\bf A}_{\rm T\varphi}({\bf 0})\right]^2+U_{\rm C\varphi}({\bf r}) + H_{\rm ph}. \label{ceda}   
\end{align}
The condition $\lambda_0 \gg a_0$, presupposed by the EDA, also implies that modes with $k\geq a_0^{-1}$ should not become significantly populated, in which case one can reduce the cut-off $k_\varphi$ within the light-matter interaction to an inverse atomic size. Replacing in Eq.~(\ref{ceda}) ${\bf A}_{\rm T\varphi}({\bf 0})$ with ${\bf A}_{\rm T\ell}({\bf 0})$ where $k_\ell \sim a_0^{-1}$, will then yield an effectively equivalent dipolar Hamiltonian that we denote $H_{\rm C}^\ell$.

To better justify this procedure let us decompose the classical transverse potential as
\begin{align}
{\bf A}_{\rm T\varphi}(t,{\bf r}(t)) = \int d^3 k \sum_\lambda {\bf A}_{\rm T\varphi}^{{\bf k}\lambda}(t,{\bf r}(t)) +{\rm H.c.}
\end{align}
Assuming that ${\bf A}_{\rm T\varphi}^{{\bf k}\lambda}(0,{\bf r}(0))={\bf 0}$, it is straightforward to show that for $t>0$
\begin{align}
{\bf A}_{\rm T\varphi}^{{\bf k}\lambda}(t,{\bf r}(t)) =& {i\varphi({\bf k})^2\over 2 k (2\pi)^3} \int_0^t dt' {\bf e}_\lambda({\bf k})\cdot {\dot {\bf d}}(t')e^{-i k(t-t')}\nonumber \\ &\times e^{-i{\bf k}\cdot [{\bf r}(t')-{\bf r}(t)]}.  
\end{align}
The EDA constitutes retaining only the zeroth order term in the expansion of $e^{-i{\bf k}\cdot [{\bf r}(t')-{\bf r}(t)]}$. For $k \geq a_0^{-1}$ the resulting dipole order term ${\bf A}_{\rm T\varphi}^{{\bf k}\lambda,0}(t,{\bf r}(t))$ satisfies
\begin{align}
|{\bf A}_{\rm T\varphi}^{{\bf k}\lambda,0}(t,{\bf r}(t))| \leq  \left|{a_0\varphi({\bf k})^2\over 2 (2\pi)^3} \int_0^t dt' {\bf e}_\lambda({\bf k})\cdot {\dot {\bf d}}(t')e^{-i k(t-t')}\right|.  \label{ineq}
\end{align}
The next lowest order term, which is neglected in the EDA, is
\begin{align}
{\bf A}_{\rm T\varphi}^{{\bf k}\lambda,1}(t,{\bf r}(t)) =& {\varphi({\bf k})^2\over 2 (2\pi)^3} \int_0^t dt' {\bf e}_\lambda({\bf k})\cdot {\dot {\bf d}}(t')e^{-i k(t-t')}\nonumber \\ &\times {\hat {\bf k}}\cdot [{\bf r}(t')-{\bf r}(t)].
\end{align}
Noting that $|{\hat {\bf k}}\cdot [{\bf r}(t')-{\bf r}(t)]|\sim a_0$ we see that the terms by which the dipolar Hamiltonians $H_{\rm C}^\varphi$ and $H_{\rm C}^\ell$ differ, given by the left hand-side of inequality (\ref{ineq}), are bounded from above by a quantity of the same order as terms already neglected within the EDA. The two Hamiltonians must therefore be considered equivalent, at least for describing {\em real} emission and absorption processes.

Even within the EDA, one should expect high-energy ($k>k_\ell$) {\em virtual} photons to occur, at least transiently. To test this, we let $F$ denote a general form factor and consider the Lamb shift of the $2s$ level. Second-order perturbation theory using $H_{\rm C}^F$, including mass renormalisation, eventually yields as the measurable shift \cite{craig_molecular_1998}
\begin{align}
\Delta_F \epsilon_{2s} = -{2\alpha_{\rm fs} \over 3\pi m^2}\sum_n \omega_{n2s}|{\bf p}_{n2s}|^2\int_0^\infty dk\, F(k)^2 {k_{n2s}\over k_{n2s}+k}\label{ls}
\end{align}
where $m$ is to be interpreted as the observed mass, and $\omega_{n2s}=\epsilon_n-\epsilon_{2s}$. Assuming a hard cut-off $F(k) = 1-\theta(k-k_F)$ with $k_F\gg k_{2s}$, one obtains the famous Bethe-log result
\begin{align}
\Delta_F \epsilon_{2s} = {\alpha^5 m \over 6\pi}\beta_{2s}^F,
\end{align}
where
\begin{align}
\beta_{2s}^F = \ln {k_F \over {\bar k}_{2s}}.
\end{align}
Here ${\bar k}_{2s}$ is an average energy difference between the $2s$ level and all others. Letting $\tau = k_\ell/ k_\varphi$ and $\sigma = {\bar k}_{2s} /k_\varphi$, the ratio of the shifts computed using cut-offs $k_\varphi$ and $k_\ell$ is
\begin{align}
{\cal R}_{2s} :={\Delta_\varphi \epsilon_{2s} \over \Delta_\ell \epsilon_{2s}} = {\beta_{2s}^\varphi \over \beta_{2s}^\ell} = -{\ln \sigma \over \ln \tau -\ln \sigma}.\label{ratioshift}
\end{align}
If we choose $k_\varphi \sim m$ as a natural non-relativistic cut-off and $k_\ell \sim a_0^{-1} = \alpha_{\rm fs} k_\varphi$, then $\tau =\alpha_{\rm fs}$. Following Bethe by using the value $\sigma = 8.9\alpha_{\rm fs}^2$, then yields the ratio ${\cal R}_{2s} \approx 2.7$, such that $\Delta_\ell \epsilon_{2s}$ is significantly different from the measured value, which is close to $\Delta_\varphi \epsilon_{2s}$. Apparently then, virtual photon effects can be well-described within the EDA, but this will generally require a relativistic cut-off $k_\varphi \sim m$.

We note before continuing that a standard scaling argument shows that the Coulomb gauge interaction is indeed ``weak" due to the smallness of $\alpha_{\rm fs}$ \cite{spohn_dynamics_2004}, legitimising the use of second-order perturbation theory. The interaction is a sum of order $q$ and order $q^2$ parts $V_{\rm C}^F=V_{\rm C}^{F1}+V_{\rm C}^{F2}$, and since $pA_{\rm T}/p^2 = A_{\rm T}^2/(pA_{\rm T})$ the ratio $V_{\rm C}^{F1}/V_{\rm C}^{F2}$ is of the same order as the ratio $H_m/V_{\rm C}^{F1}$ \cite{cohen-tannoudji_photons_1989}. It follows that $V_{\rm C}^{F1}$ and $V_{\rm C}^{F2}$ are also {\em individually} weak.

%\begin{itemize}
%\item Give two LWL Coulomb gauge Hamiltonians $H_0^l$ and $H_0^M$ cut off at $k_l\sim a^{-1}$ and $k_\varphi \sim \lambda_c^{-1}$ respectively. State the conditions under which they are equivalent (modes $k>k_l$ are never populated).
%\item Show that dipole contributions for $k>a^{-1}$ in $H_0^M$ are comparable to higher order multipole terms for $k<a^{-1}$ retained in both Hamiltonians. 
%\item Show that the individual terms in the interaction are each weak perturbations of the standard unperturbed Hamiltonian.
%\end{itemize}

\subsubsection{Dipole gauge}

A dipole gauge Hamiltonian $H_{\rm M}^F$ can be defined as $H_{\rm M}^F = R_F H_{\rm C}^F R_F^\dagger$ where $F=\varphi,\,\ell$ and the required Power-Zienau-Woolley (PZW) transformation in the EDA is   
\begin{align}
R_F = \exp\left[-i {\bf d}\cdot {\bf A}_{{\rm T}F}({\bf 0})\right],
\end{align}
which displaces ${\bf p}$ by $q{\bf A}_{{\rm T}F}({\bf 0})$ and ${\bf \Pi}$ by the dipolar transverse polarisation
\begin{align}
{\bf P}_{{\rm M}_F\rm T}({\bf x}) = {\bf d}\cdot \delta_F^{\rm T}({\bf x}).
\end{align}
Thus,
\begin{align}
H_{\rm M}^F = H_m+ {1\over 2}\int d^3 x \left(\left[{\bf \Pi}({\bf x})+{\bf P}_{{\rm M}_F\rm T}({\bf x})\right]^2+{\bf B}({\bf x})^2\right).\label{HmF}
\end{align}
Note that if one applies $R_\ell$ to $H_{\rm C}^\varphi$ as $R_\ell H_{\rm C}^\varphi R_\ell^\dagger$ then one obtains $H_{\rm M}^\ell$  plus additional transverse potential-dependent interaction terms, because in this case the potential is not completely eliminated from the mechanical momentum, which instead reads
\begin{align}
m{\dot {\bf r}} = {\bf p} -q\left[{\bf A}_{\rm T\varphi}({\bf 0})-{\bf A}_{\rm T\ell}({\bf 0})\right].
\end{align}
The potential contribution here is restricted to modes $k\geq k_\ell \gg \lambda_0$, which we assumed in the Coulomb gauge do not become appreciably populated. This shows directly that the two dipole gauge Hamiltonians, $H_{\rm M}^F,~F=\varphi,\,\ell$, are indeed equivalent under the same conditions that the corresponding Coulomb gauge Hamiltonians are equivalent. 

The same $S$-matrix elements are obtained in the two gauges if and only if $H_m$ is defined in the same way in both gauges \cite{woolley_gauge_1998,woolley_charged_1999}, with the standard definition being given by Eq.~(\ref{hm1}). In this case, the polarisation self-term must be included within the interaction, which is then the sum of order $q$ and order $q^2$ terms;
\begin{align}
        V_{\rm M}^F =   {\bf d}\cdot {\bf \Pi}_F({\bf x}) +\delta U_{{\rm M}_F} \equiv V_{\rm M}^{F1} + V_{\rm M}^{F2}. \label{dgint}
\end{align}
If $V_{\rm C}^F$ is a weak perturbation of $h=H_m+H_{\rm ph}$ then it must be possible to consider $V_{\rm M}^F$ a weak perturbation too. We have seen however, that while $V_{\rm M}^{\ell2} = \delta U_{{M_\ell}}$ is weak, $V_{\rm M}^{\varphi2}=\delta U_{\rm M_\varphi}$ is not. It follows that in the case $F=\ell$, the individual terms in Eq.~(\ref{dgint}) are separately weak, whereas in the case $F=\varphi$, the individual terms are not separately weak, even though their combination must be. Thus, requiring a weak perturbation in the multipolar gauge does not imply that $\delta U_{{\rm M}_F}$ must be included within $H_m$ nor that the cut-off be reduced down to $k_\ell \lesssim a_0^{-1}$ if it is included within the interaction. Condition I is a stronger condition than the weakness of the interaction. It requires that every Hamiltonian term, additional to the standard unperturbed Hamiltonian, is {\em separately} weak.

A simple (crude) adiabatic elimination argument provides additional insight. We begin by distinguishing the Coulomb and dipole gauge photon operators using labels ${\rm C}$ and ${\rm M}$, and note that the two sets are related by
\begin{align}
    a_{\rm M\lambda}(t,{\bf k}) &= R_\varphi^\dagger a_{\rm C\lambda}(t,{\bf k}) R_\varphi \nonumber \\ &= a_{\rm C\lambda}(t,{\bf k}) - {i\varphi({\bf k}) \over \sqrt{2\omega (2\pi)^3}} {\bf e}_\lambda({\bf k})\cdot {\bf d}(t).
\end{align}
Suppose, following our analysis in Sec.~\ref{couleda}, that in the Coulomb gauge, modes with $k\geq k_\ell$ are never appreciably populated, such that $a_{\rm C\lambda}(t,{\bf k}) = 0$, then
\begin{align}
    a_{\rm M\lambda}(t,{\bf k}) &= -{i\varphi({\bf k}) \over \sqrt{2\omega (2\pi)^3}} {\bf e}_\lambda({\bf k})\cdot {\bf d}(t).\label{AE}
\end{align}
This is equivalent to assuming that in the dipole gauge modes with $k\geq k_\ell$ do not {\em vary} appreciably, that is, Eq.~(\ref{AE}) holds if and only if ${\dot a}_{\rm M\lambda}(t,{\bf k}) = -i[ a_{\rm M\lambda}(t,{\bf k}),H_{\rm M}^\varphi ] =0$ \cite{de_bernardis_cavity_2018}. Decomposing ${\bf \Pi}_\varphi (t,{\bf 0})$ and $\delta U_{{\rm M}_\varphi}$ as
\begin{align}
{\bf \Pi}_\varphi (t,{\bf 0}) &= \int d^3 k \sum_\lambda {\bf \Pi}_{\varphi \lambda} (t,{\bf k}), \\
\delta U_{{\rm M}_\varphi}(t) &= \int d^3 k \sum_\lambda \delta U_{{\rm M}_\varphi
\lambda}(t,{\bf k})
\end{align}
and using Eq.~(\ref{AE}) gives
\begin{align}
  &V_{\rm M \lambda}^{\varphi 1}(t,{\bf k}) 
  := {\bf d}(t)\cdot {\bf \Pi}_{\varphi \lambda} (t,{\bf k})=  -{\varphi({\bf k})^2 \over (2\pi)^3} [{\bf e}_\lambda({\bf k})\cdot {\bf d}(t)]^2, \\
&{1\over 2}{\bf \Pi}_{\varphi \lambda} (t,{\bf k})^2 = {\varphi({\bf k})^2 \over 2(2\pi)^3} [{\bf e}_\lambda({\bf k})\cdot {\bf d}(t)]^2 = 
  \delta U_{{\rm M}_\varphi
\lambda}(t,{\bf k}).  
\end{align}
Thus, the always non-negative transverse electric energy density
\begin{align}
{1\over 2}{\bf E}_{\rm T\varphi}(t,{\bf k})^2 = {1\over 2}{\bf \Pi}_{\varphi \lambda} (t,{\bf k})^2+V_{\rm M \lambda}^{\varphi 1}(t,{\bf k})+\delta U_{{\rm M}_\varphi
\lambda}(t,{\bf k}),
\end{align}
vanishes for modes with $k\geq k_\ell$. The contribution of $V_{\rm M}^{\varphi 1}$ exactly cancels the sum of contributions from $\delta U_{{\rm M}_\varphi}$ and the non-magnetic part of $H_{\rm ph}$. We conclude that the Hamiltonian $H_{\rm M}^\varphi$ does not receive any contribution from modes with $k\geq k_\ell$, despite that these modes contribute to both $V_{\rm M}^{\varphi 1}$ and $V_{\rm M}^{\varphi 2}$, implying that neither term is individually weak.

We finally comment on the Lamb shift calculated within the dipole gauge. We have seen via calculation in the Coulomb gauge that for the $2s$ level, the shift is strongly dependent on the cut-off $k_F$. The same on-shell matrix element is obtained in the dipole-gauge provided the same (standard) definition of $H_m$ is employed. In this calculation, one observes the direct cancellation of the contribution $\bra{2s}\delta U_{\rm M_\ell}\ket{2s}$, by a contribution from $V^{F1}_{\rm M}$. The cancelled term is of the same order as the final Lamb shift result \cite{vukics_fundamental_2015}. After further removal of a material state-independent term and mass renormalisation, Eq.~(\ref{ls}) remains. If, on the other hand, $\delta U_{\rm M_\ell}$ is included within $H_m$, then the shift must be recomputed.
We might expect the result to differ significantly from the measured value, because the calculation could no longer feature the same cancellation of the contribution of $\delta U_{\rm M_\ell}=V_{{\rm M}_\varphi}^2$ by a term coming from $V_{{\rm M}_\varphi}^1$.

\subsection{Beyond the EDA}\label{Beyond the EDA}

We give a brief indication of how avoiding the EDA modifies single-atom calculations. The natural cut-off to assume is at the onset of relativistic energies $k_\varphi \sim m$. In the case of real emission and absorption, the on-shell condition is satisfied, so for low-energy transitions ${\bf k}\cdot {\bf r} \sim a_0 k \sim \alpha_{\rm fs}$. Thus, corrections to the EDA should be small.

Simple expressions are straightforwardly obtainable if one considers a two-level atom. In the case of hydrogen, restricting one's attention to the principal levels $1s$ and $2p$ reduces the material Hilbert space to the span of four basis states, $\{\ket{g},\ket{e_m},~m=-1,0,1\}$. We denote the projection onto this space by $P$ and replace the canonical operators $r$ and $p$ by their projections $PrP$ and $PpP$ within $V_g$. The spontaneous emission rate for the transition $2p\to 1s$ with $m=0$ is readily found to be
\begin{align}
{\tilde\Gamma}_{2p1s} = \Gamma_{2p1s} C(k_{2p1s}r_{2p1s})
\end{align}
where $\Gamma_{2p1s}$ is the standard rate obtained within the EDA and $C$ is a correcting factor defined by
\begin{align}
C(\mu) =\frac{1}{2}-\frac{3 \cos (2 \mu)}{8 \mu^2}+\frac{3 \sin (2 \mu)}{16 \mu^3}.
\end{align}
The EDA result with $C=1$ is recovered by letting the argument of $C$ tend to zero, equivalently, by expanding $C$ to the first non-zero term. Deviations of $C$ from unity by more than $10^{-3}$ require $\mu >0.05$. For atomic hydrogen $\mu = k_{2p1s}r_{2p1s} \sim 0.002$ where $r_{2p1s}$ is the transition moment for the $m=0$ state. As expected, the contribution of the lowest transition to the deviation from the EDA result is minuscule. The largest non-EDA contributions are expected to come from high-energy transitions. Using $V=-q{\bf A}_{\rm T}({\bf r})\cdot {\bf p}/m$, one immediately sees that the rate is determined by the matrix element
\begin{align}
    {\bf M}&=\bra{2p}e^{i{\bf k}\cdot {\bf r}}{\bf p}\ket{2s}
\end{align} 
evaluated on-shell; $k=k_0:= \omega_{2p}-\omega_{1s}$. We therfore let $e^{i{\bf k}\cdot {\bf r}}=1+i{\bf k}\cdot {\bf r}-({\bf k}\cdot {\bf r})^2/2$ and note that due to angular symmetry the linear term $i{\bf k}\cdot {\bf r}$ does not contribute to the rate. We let ${\bf p}=-i\nabla$, note that the $2s$ wave-function is purely radial, and note that the angular integrals within
\begin{align}
    {\bf M}_{\rm dip}&=\bra{2p}{\bf p}\ket{2s}\\
    {\bf M}_{\rm quad}&=-{1\over 2}\bra{2p}({\bf k}\cdot {\bf r})^2{\bf p}\ket{2s}
\end{align}
do not affect their parametric scaling. The relative sizes are therefore determined entirely by the radial integrals
\begin{align}
    I_{\rm dip} = \int_0^\infty dr \, r^2 R_{2p}(r) \frac{dR_{1s}}{dr}, 
\end{align}
and
\begin{align}
I_{\rm quad} = k_0^2 \int_0^\infty dr \, r^4 R_{2p}(r) \frac{dR_{1s}}{dr}
\end{align}
with ratio
\[
\frac{I_{\rm quad}}{I_{\rm dip}} = \frac{80}{9} \, a_0^2 k_0^2 \sim \alpha^2.
\]
Thus, the leading correction to the $2p\to 1s$ emission rate beyond dipole order is minuscule, of order $\alpha^4$.

\section{Two atom interactions}

One of the primary advantages of multipolar QED is its relative localisation of material subsystems, whose interactions are then mediated entirely via causal local fields. Since regularisation inevitably compromises spatial localisation, we now consider interactions between the atoms at ${\bf 0}$ and ${\bf R}=R{\hat {\bf x}}$. 

\subsection{Direct material interactions in the EDA}\label{2 atom direct interaction section}

We begin by considering the direct atom-atom interaction [Eq.~(\ref{Ug12})]. Throughout we assume a Lorentzian cutoff function $F$ with cut-off $k_F$. We denote by $\varphi$ and $\ell$ the particular cases $k_F=k_\varphi \sim \lambda_c^{-1}$ (relativistic cut-off) and $k_F=k_\ell\sim a_0^{-1}$ (dipolar cut-off).

\subsubsection{Coulomb gauge}\label{couldirect}

In the Coulomb gauge (${\bf g}_{\rm T}=0$), Eq.~(\ref{Ug12}) is simply the overlap of the electrostatic fields of the two atoms
\begin{align}
U_{\rm C_F 12} &= \int d^3 x\, {\bf E}_{F {\rm L}1}({\bf x}) \cdot {\bf E}_{F {\rm L}2}({\bf x}) \nonumber \\ &= \int d^3 x d^3 x' {{\rho_{F1}({\bf x})\rho_{F2}({\bf x}')} \over 4\pi|{\bf x}-{\bf x}'|}.
\end{align}
For a relativistic cut-off $k_M \sim \lambda_c^{-1}$ this interaction is essentially indistinguishable from the bar electrostatic interaction.

In the EDA one obtains the regularised dipole-dipole interaction
\begin{align}\label{u12c}
U_{\rm C_F 12} =& {\bf d}_1 \cdot \delta_{F^2}^{\rm L}({\bf R})\cdot {\bf d}_2 %\nonumber\\ =&{\bf d}_1 \cdot \delta^{\rm L}({\bf R})\cdot {\bf d}_2\left[1-f(k_\varphi R)\right] \nonumber \\ &+ {({\bf d}_1\cdot {\hat {\bf R}})({\bf d}_2\cdot {\hat {\bf R}}) \over 4\pi R^3}g(k_\varphi R)
\end{align}
and the Hamiltonian reads
\begin{align}
     H_C^F =\sum_{n = 1,2} \left(H_{m,n}+V_{Cn}\right)+U_{C_F 12}+H_{\rm ph}\label{hcdip}
\end{align}
in which $V_{Cn}$ is the light-matter interaction of dipole $n$.

In the EDA, it is assumed that modes with $k>k_\ell \sim a_0^{-1}$ do not contribute to real processes, in which case the relativistic cut-off can be reduced to $k_\ell \sim a_0^{-1}$ within the light-matter coupling (see Sec. \ref{couleda}). If $U_{C_F 12}$ and the light-matter coupling  possess different cut-offs, then the physically expected cancellation of instantaneous terms cannot completely occur. Thus, to obtain properly retarded $S$-matrix elements, one must also reduce the cut-off within the dipole-dipole interaction, giving $U_{C_\ell 12}$, and corresponding Hamiltonian $H_C^\ell$.
%We therefore restrict our attention to the Hamiltonian $H_C^F$.

For ${\bf d}_1$ and ${\bf d}_2$ along the $z$-direction, as occurs within the static energy transfer matrix element $\bra{2p,1s}U_{\rm C_F 12}\ket{1s,2p}$ with $m=0$, the ratio $U_{\rm C_F 12}/U_{\rm C 12}$, where $U_{\rm C 12}$ denotes the $k_F \to \infty$ limit, is
\begin{align}
\theta = {U_{\rm C_F 12}\over U_{\rm C12}}= {\delta_{F^2 zz}^{\rm L}({\bf R}) \over \delta_{zz}^{\rm L}({\bf R})} = 1-e^{-\mu}\left(1+\mu+{1\over 2}\mu^2 %+\frac{1}{4}\mu^{3}
\right).\label{ratioc}
\end{align}
where $\mu= k_FR$. Deviation from unity by more than $10^{-3}$ requires $\mu <15$ with a deviation of roughly $0.003$ when $\mu=10$. Thus, for $k_F = k_\ell$ significant deviations begin to occur for separations $R < 10a_0$, growing exponentially as $R$ is decreased. Here modes with $k>a_0^{-1}$, for which the EDA is not generally valid, start to become important (Fig.~\ref{fig:ratios}).

In the generic problem of energy transfer (Sec.~\ref{sec:res}), the static interaction Hamiltonian only contributes to the coherent inter-dipole interaction, producing an energy shift analogous to the single-dipole Lamb shift, for which we have seen that modes $k>k_\ell$ are indeed important. We have also seen, however, that a sensible dipolar Lamb shift is obtained if (and only if) one chooses $k_F =k_\varphi\sim 100a_0^{-1}$. By choosing this $k_F$ within the two-dipole Hamiltonian, one sees that $\theta$ deviates from unity only when $R<a_0/10$ (Fig.~\ref{fig:ratios}). In particular, $\theta =1$ for all separations above the limit of atomic overlap. 

%Nevertheless, since there are no quadrupole contributions connecting the $n=2$ and $n=1$ principal levels, the corrections to the EDA result promise to remain small.CAN WE COMPUTE THE EXACT ELECTROSTATIC INTERACTION NUMERICALLY FOR SMALL $R$? RESULTS SHOULD BE INCLUDED IN THE BEYEND THE EDA SECTION BELOW.

\subsubsection{Dipole gauge}
For the multipolar choice in Eq.~(\ref{convG}) the contributions of the longitudinal and transverse polarisations are independently regularised via $\varphi$ and $\ell$ respectively [cf.~Eq.~(\ref{Upzw})]. This gives the total direct material interaction
\begin{align}\label{u12full}
U_{{\rm M}_{\varphi \ell}12} = U_{C_\varphi 12}+\delta U_{M_\ell 12}
\end{align}
where
\begin{align}
\delta U_{M_\ell 12}:=\int d^3 x \,{\bf P}_{\ell \rm T,1}({\bf x})\cdot {\bf P}_{\ell \rm T,2}({\bf x}).
\end{align}
If $k_\ell = k_\varphi$ and we take the point limit, $k_\varphi \to \infty$, then $U_{{\rm M}_{\varphi \ell}12}$ is strictly zero for non-overlapping atoms. If instead $k_\ell = k_\varphi = k_F$, but the value is kept finite one obtains exponentially localised atomic polarisations, such that
\begin{align}\label{UMfull}
U_{M_F 12}:=\int d^3 x \,{\bf P}_{F,1}({\bf x})\cdot {\bf P}_{F,2}({\bf x})
\end{align}
is exponentially suppressed as the interatomic separation $R$ increases.

In the EDA, the multipolar Hamiltonian is related to $H_C^F$ in Eq.~(\ref{hcdip}) by $H_M^F= R_F H_C^F R_F^{-1}$ where $R_F= {\exp}\left[-i{\bf d}_1\cdot {\bf A}_{{\rm T}F}({\bf 0}) -i {\bf d}_2 \cdot {\bf A}_{{\rm T}F}({\bf R})\right]$. It can be partitioned in the same way as $H_{\rm C}^F$ as
\begin{align}
     H_M^F =\sum_{n = 1,2} \left(H_{m,n}+V_{Mn}\right)+U_{M_F 12}+H_{\rm ph}
\end{align}
where
\begin{align}
U_{{\rm M}_F12} = & {\bf d}_1 \cdot {\bf d}_2 \delta_{F^2}({\bf R}).\label{u12}
\end{align}
is the dipole approximation of Eq.~(\ref{UMfull}). Supposing again that ${\bf d}_1$ and ${\bf d}_2$ point in the $z$-direction, the ratio
\begin{align}
\phi = {U_{{\rm M}_F 12} \over U_{C_F12}} = 1+ {\delta_{F^2zz}^{\rm T}({\bf R}) \over \delta_{F^2zz}^{\rm L}({\bf R})}=\frac{\mu ^3 e^{-\mu }}{2-2 e^{-\mu } \left[1+\left(1+\frac{1}{2}\mu\right)\right]}
\end{align}
compares the direct interaction terms within dipolar Hamiltonians $H_C^F$ and $H_M^F$.

\begin{figure}[h]
\centering
\includegraphics[width=\linewidth]{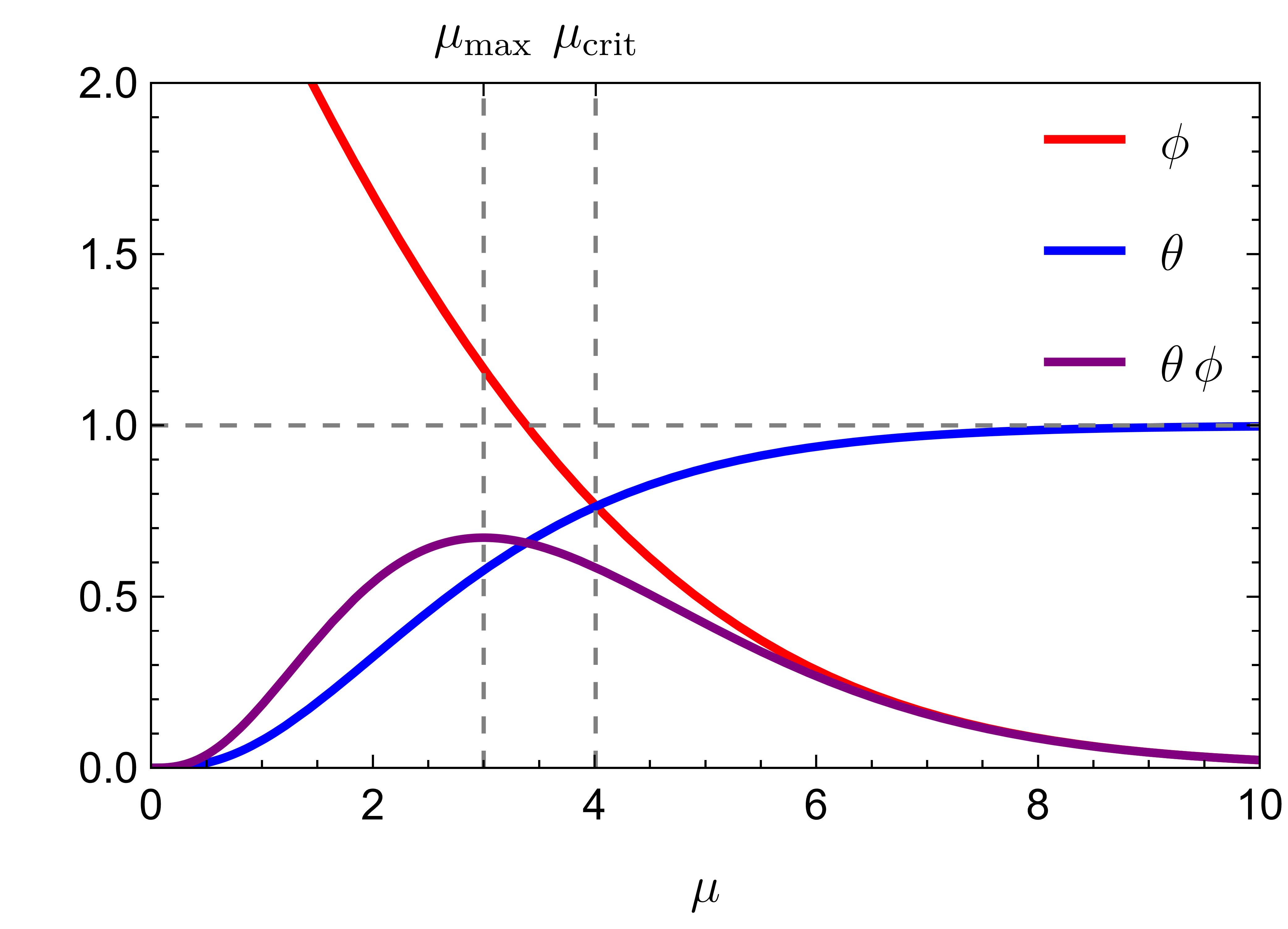}
\caption{%The ratios $\phi$ (blue),  $\theta$ (green), and $\theta\phi ={1\over 2}\mu^3e^{-\mu}$ (gold) are plotted with $\mu=k_F R$. As the effect of regularisation becomes appreciable $\theta$ deviates from unity and $\theta\phi$ deviates from zero. The blue curve reveals, for each value of $k_F$, which values of $R$ result in a dipole-gauge interaction $U_{M_F12}$ that is comparable to the corresponding Coulomb gauge (electrostatic) interaction $U_{C_F12}$. The point at which the two are equal is $\mu_{\rm crit}\approx 3.834$. The gold curve similarly compares the multipolar and bare ($k_F\to \infty$) Coulomb interactions. Its maximum value is $0.672$ occurring at $\mu=3$.
The ratios $\phi$ (red),  $\theta$ (blue), and $\theta\phi ={1\over 2}\mu^3e^{-\mu}$ (purple) are plotted with $\mu=k_F R$. As the effect of regularisation becomes appreciable $\theta$ deviates from unity and $\theta\phi$ deviates from zero. The red curve reveals, for each value of $k_F$, which values of $R$ result in a dipole-gauge interaction $U_{M_F12}$ that is comparable to the corresponding Coulomb gauge (electrostatic) interaction $U_{C_F12}$. The point at which the two are equal is $\mu_{\rm crit}\approx 4.01$. The purple curve similarly compares the multipolar and bare ($k_F\to \infty$) Coulomb interactions. Its maximum value is $0.672$ occurring at $\mu=3$.}
\label{fig:ratios}
\end{figure}

The ratio $\phi$ becomes appreciably non-zero precisely when the ratio in Eq.~(\ref{ratioc}) deviates significantly from unity, i.e.\ for $\mu \lesssim 10$. For $k_F = k_\ell \sim a_0^{-1}$ this occurs for separations $R \lesssim 10 a_0$, which is exactly the regime where modes with $k > k_\ell$ begin to matter. If $k_F = 1/(n a_0^{-1})$ with $n = 0,1,2,\ldots$, then $\phi = 1$ when $R = \mu_{\mathrm{crit}}\, n a_0$, where $\mu_{\mathrm{crit}} \approx 4.01$ %$\mu_{\mathrm{crit}} \approx 3.834$
(see Fig.~\ref{fig:ratios}). For example, if $k_F = 0.5 a_0^{-1}$, then at $R = 8.02 a_0$ the multipolar direct interaction already equals the Coulomb-gauge direct interaction, and becomes larger for smaller separations. In this sense, employing a low cut-off severely compromises the removal of direct material interactions that is usually regarded as characteristic of the multipolar theory. There is an increasingly large lower bound on the separations $R$ for which this property holds as $k_F$ is reduced. Modes in the range $k_\ell \le k \le k_\phi$ are evidently needed to increase localisation of the multipolar polarisation. Indeed, for $k_F = k_\phi \sim 100 a_0^{-1}$, the ratio $\phi$ is negligible for all separations. The exponential dependence ensures that even modest increases of the cut-off above $k_F \sim a_0^{-1}$ suppress $\phi$ rapidly; for instance, if $k_F = 5 a_0^{-1}$, then $\phi \approx 0.02$ at $R = 2 a_0$. %Allowing different cut-offs $k_\ell < k_\phi$ in Eq.~(\ref{u12full}) leads to a more involved dipolar approximation: only modes with $k \le k_\ell$ are suppressed, while the Coulomb interaction from modes with $k_\ell < k \le k_\phi$ remains.
%EXPLAIN WHAT THIS GRAPH SHOWS. FOR $k_\ell$ LARGE $\sim m$ THE DIPOLE GAUGE POLARISATION IS HIGHLY LOCALISED, AND WE SHOULD SEE THAT $\delta_{\ell^2zz}^{\rm T} \approx \delta_{zz}^{\rm T} = -\delta_{zz}^{\rm L}$ FOR $R\neq 0$, SO THAT THE RATIO ABOVE IS APPROXIMATELY ZERO. FOR WHAT VALUES OF $\mu$ IS THE RATIO OF ORDER UNITY? IN THIS CASE THE LOCALISATION OF THE DIPOLE GAUGE POLARISATION IS BECOMING COMPARABLE TO THAT OF THE ELECTROSTATIC FIELD OF A DIPOLE (THE COULOMB GAUGE POLARISATION). THUS, HOW MUCH MORE LOCALISED IS THE DIPOLE GAUGE POLARISATION THAN THE COULOMB GAUGE ONE FOR GIVEN CUTOFFS $k_\ell$?.

\subsection{Resonant energy transfer in the EDA}\label{sec:res}

The second-order matrix element for resonance energy transfer between two dipoles is defined by
\begin{align}\label{tfi}
T=&\bra{i}V\ket{f}+\sum_{I}\frac{\bra{i}V\ket{I}\bra{I}V\ket{f}}{E_{i}-E_{I}},\nonumber \\
\equiv& U_{12}+T_{\rm lm}.
\end{align}
Where the initial state  $\ket{i}=\ket{e,g,0}$ consists of dipole 1 excited and dipole 2 de-excited with no photons, while in the final state all (bare) energy has been transferred to dipole 2; $\ket{f}=\ket{g,e,0}$. Only the light-matter interaction contributes to the second term, while only the direct material interaction contributes to the first term. The real part of $T$ is interpretable as a joint shift, while the imaginary part describes decay. The direct interaction term $U_{12}$ contributes to the shift only.

The matrix element $T$ is gauge non-relative, meaning that it does not depend on the choice of interaction Hamiltonian $V$, despite the implicit gauge-relativity of the bare vectors $\ket{i}$ and $\ket{f}$. Specifically, the bare vector $\ket{n}$
represents a different physical state in different gauges whenever the light-matter coupling is finite.

The Coulomb gauge photonic subsystem implicitly includes a static contribution characterised by singular behaviour at zero mode frequency. The calculation of the complete energy transfer matrix element in the standard limit $k_F\to \infty$ in this gauge reveals a term coming from the light-matter interaction [2nd term in Eq.~(\ref{tfi})] that exactly cancels the direct Coulomb interaction [1st term in Eq.~(\ref{tfi})] to give the expected fully retarded result \cite{craig_molecular_1998}.

The multipolar gauge is designed to achieve this cancellation at the operator level by removing the static singularity from the Coulomb gauge photonic subsystem and incorporating these degrees of freedom into the material subsystem. The Coulomb field outside the atom is thereby exactly cancelled, so that the 1st term in Eq.~(\ref{tfi}) vanishes, and a local causal field mediates all interactions between the resulting material systems, via the 2nd term in Eq.~(\ref{tfi}).
\begin{figure}[t]
\centering
\includegraphics[width=\linewidth]{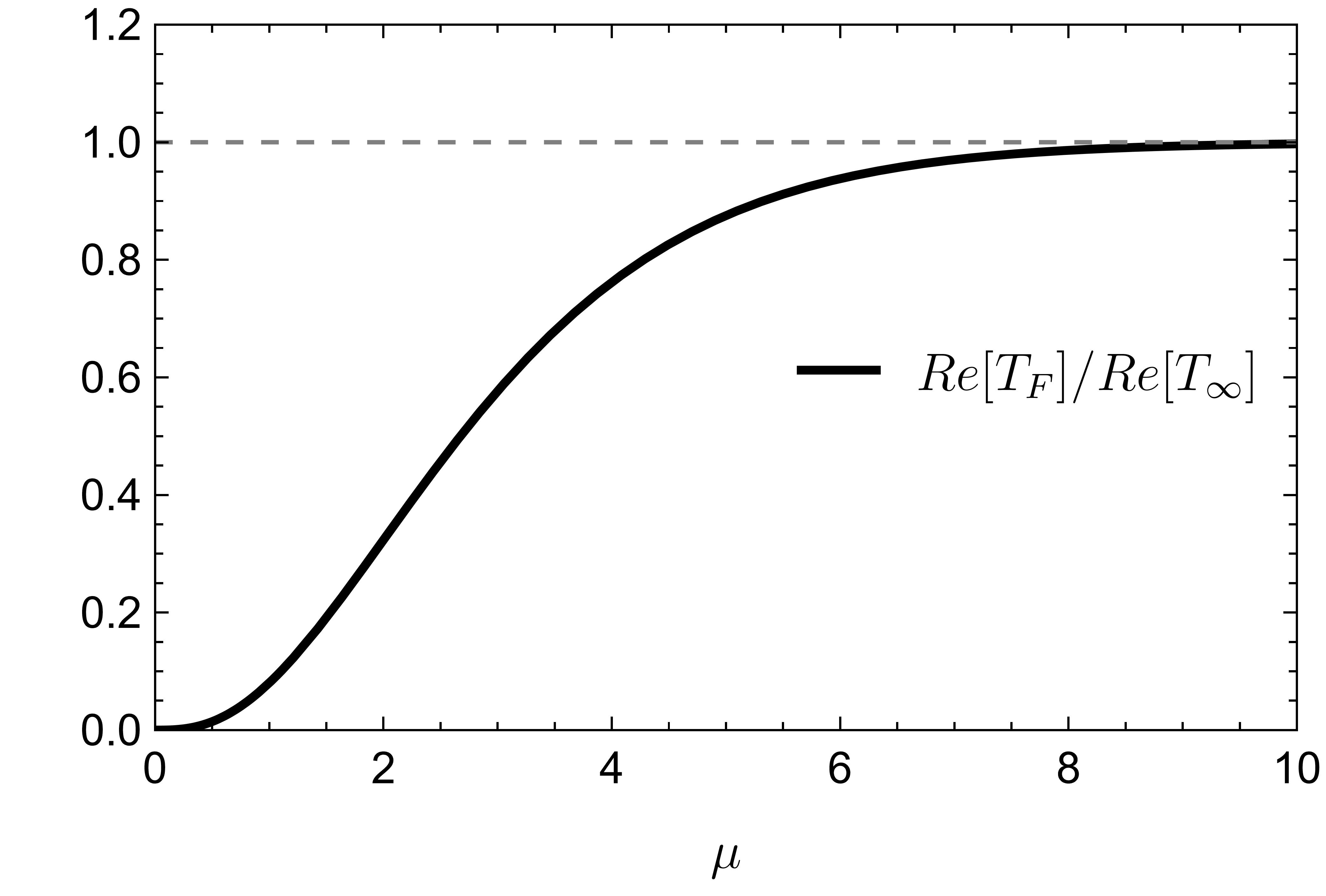}
\caption{The ratio of the regularised transition element to the point-charge ($k_{F}\rightarrow \infty$) transition element against $\mu$. $\eta=k_{F}/k_{eg}$ is held fixed at $\eta\equiv100$. For a $k_{F}\sim a^{-1}_0$ we have agreement with the point-case when $R\sim10[a_0]$, whereas for $k_F \sim \lambda_{\rm C}^{-1}$ we have ${\rm Re}[T]_{F}]/{\rm Re}[T_{\infty}]\approx1$ for all meaningful $R$.}
\label{fig:Tratio}
\end{figure}

Restricting our attention to regularised Coulomb and dipole gauges, the matrix element becomes
\begin{align}\label{partitionT}
T_F &= U_{{\rm C}_F 12}+T_{\rm C_F,lm} \nonumber \\
&= U_{{\rm M}_F 12}+T_{\rm M_F,lm}
\end{align}
where the direct interaction terms $U_{{\rm C}_F 12}$ and $U_{{\rm M}_F 12}$ were compared in Sec.~\ref{2 atom direct interaction section}. Since the imaginary part of $T$ is independent of the direct interaction and essentially independent of the cut-off, we focus on the real part. Assuming again identical dipoles, with ${\bf d}_1={\bf d}_2 = d{\hat {\bf z}}$, and ${\bf R}=R{\hat {\bf x}}$, one obtains the standard result
\begin{align}
\beta{\rm Re}[T_\infty] = \frac{\sin (\nu)}{\nu^2}+\frac{\cos (\nu)}{\nu^3}-\frac{\cos (\nu)}{\nu}\label{Tinf}
\end{align}
where $\nu = k_{eg} R$ and $\omega_{eg} = ck_{eg}$ is the $e\to g$ transition frequency. For convenience we have defined the proportionality constant $\beta = 4\pi /(d^2 k_{eg}^3)$.  Since $T_F$ is invariant, the difference in light-matter terms is simply the negative of the difference in direct interaction terms. In the multipolar gauge ${\rm Re}[T_\infty]={\rm Re}[T_{\rm M_\infty, lm}]$, that is, only the light-matter interaction contributes when $k_F\to \infty$, whereas in the Coulomb gauge ${\rm Re}T_\infty={\rm Re}[T_{\rm C_\infty, lm}]+U_{C12}$ and $U_{C12}$ is signifincant in the near field, $\nu \ll1$. More generally, for $k_F$ finite,
\begin{align}
\beta {\rm Re}[T_{\rm M_F,lm}]=& \frac{\mu^4}{\left(\mu^2+\nu^2\right)^2}\bigg(\frac{\sin (\nu)}{\nu^2}+\frac{\cos (\nu)}{\nu^3}-\frac{\cos (\nu)}{\nu}\nonumber \\&-\frac{e^{-\mu}}{2 \nu^3} \left[\mu^3+\mu^2+\mu \left(\nu^2+2\right)-\nu^2+2\right]\bigg).% \nonumber \\ &+\frac{e^{-\mu} \mu^3}{2 \nu^3}
\end{align}
The first term is as in Eq.~(\ref{Tinf}), but with a prefactor of $F$, while the second term is due entirely to the finiteness of $k_F$. The Coulomb gauge light-matter term is then found according to Eq.~(\ref{partitionT}) from $T_{{\rm C}_F,{\rm lm}}-T_{{\rm M}_F {\rm lm}} =U_{{\rm M}_F 12}-U_{{\rm C}_F 12}$. The relative size of this difference was determined in Sec.~\ref{2 atom direct interaction section}. The ratio ${\rm Re}[T_F]/{\rm Re}[T_\infty]$ is shown in Fig.~\ref{fig:Tratio} as a function of $\mu$ while $\eta = k_F/k_{eg} = \mu/\nu$ is held fixed at $\eta=100$. It is again seen that for $\mu\lesssim 10$ the regularisation becomes important, so for $R\lesssim 1/k_F$ modes with $k>k_F$ begin to make significant contributions. The value of the ratio is insensitive to the value of $\eta$ within the relevant regime, i.e, for all values of $\eta$ such that $k_{eg}\sim \alpha_{\rm fs}/a_0$ (see Fig.~\ref{fig:scale hierarchy}).

%Consider RET in using the full Hamiltonian:
%\begin{itemize}
%\item Determine the regime of separations for which the theory is valid (cond. 1), as the regime for which the prediction is effectively independent of the regularisation
%\item Determine the ratio of direct interaction to the matter-photon interaction in CG and DG, as a function of $k_\varphi$ and $R$. A small ratio in the DG can be identified as a desirable trait (cond. 2).
%\item Determined the regime for which DG conds. 1 and 2 are met. Note that this may typically imply a violation of cond. 0, which however, we have shown to be unnecessary.
%\end{itemize}

\subsubsection*{Discussion}

The onset of deviations from the infinite-cutoff result at $\mu = k_F R \lesssim 10$ coincides with the regime $k_F \sim a_0^{-1}$, where the field begins to resolve atomic structure, precisely the scale at which the electric–dipole approximation (EDA) ceases to be self-consistent. The sensitivity of resonant energy transfer rates and static inter-dipole interactions to $k_F$ does not signal a pathology of the regularisation procedure, but the possible breakdown of the dipolar description.

For $\mu = k_F R \lesssim 10$ a reliable description should avoid the EDA and retain modes such that $a_0^{-1}<k<\lambda_c^{-1}$. Importantly, the noteworthy physical features of the multipolar description are not inherent to the dipole approximation. In particular, regardless of the EDA, the direct interatomic interaction is exponentially suppressed, of the form $e^{-\mu}f(R)$ where $f(R)$ can be expanded in powers of $R$. Thus, if $k_F\sim \lambda_c^{-1}$, then all phenomena of the interacting theory outside of the individual atoms themselves must be interpreted as originating from the light-matter interaction, which is defined in terms of properly local and retarded transverse fields ${\bf B}$ and ${\bf D}_{\rm T}$.

Whether or not overlap effects, scaling as $e^{-R/a_0}$,  become important depends on the systems and transitions considered. A prototypical figure of merit is the Dicke critical quantity (dimensionless coupling squared) $\tau= 2\rho{d^2/\omega_0}$ where $\rho$ is the density of aligned two-level dipoles, each with transition moment $d$ and corresponding transition energy $\omega_0$. The critical value is $\tau=1$, beyond which an abnormal phase occurs. For natural atoms and a ground-state atomic transition, this occurs near the border of solidification, where atomic overlap becomes significant. The critical coupling could be achieved within a {\em dilute} gas of Rydberg atoms, but only via adjacent Rydberg levels. For natural atomic ground state transitions $d$ grows as $\omega_0$ decreases. The critical coupling would therefore require more exotic molecular states or engineered quantum emitters in which $d^2/ \omega_0$ can be made large while keeping $\rho$ constant. 

Quite generally, if $R/a_0 \gtrsim 10$, then overlap effects are essentially negligible, and the multipolar theory with vanishing direct interaction is robust. Further, for $k_F\sim a_0^{-1}$, the dipolar description is reliable when $\mu\geq 10$, which provides the same bound $R/a_0\gtrsim 10$. For $k_F$ approaching $\lambda_c^{-1}$, the two bounds are distinct and the range of permissible separations $R$ is restricted only by the condition that $R/a_0$ is sufficiently small.

\section{Conclusions}

We have developed an arbitrary-gauge regularised formulation of non-relativistic QED. Each choice of gauge defines a distinct partition of the composite system into canonical light and matter subsystems, related by unitary transformations acting on the full Hilbert space. We have analysed the consequences of Lorentzian regularisation with cut-off $k_F$ for the Coulomb and multipolar gauges. 

We have studied different partitions of the Hamiltonian into free and interaction parts by determining the purely material polarisation self-interaction, the so-called "$P^2$" term, of multipolar QED. In the electric dipole approximation, our result reduces to the known harmonic form \cite{vukics_fundamental_2015}. We have found that even for $k_F \sim a_0^{-1}$, this term can have a notable effect on perturbative physical predictions when incorporated into the free Hamiltonian. It should therefore be treated as part of the light-matter interaction Hamiltonian. Choosing $k_F \lesssim a_0^{-1}$ ensures that the individual light-matter interaction terms are separately weak. Employing such a low cut-off, however, compromises the localisation of the multipolar material charge distributions and the resulting suppression of direct material interactions, which is usually regarded as a distinct advantage of the multipolar theory. As $k_F$ is reduced, the minimum separation $R$ for which this property holds increases. Moreover, physically meaningful virtual processes responsible for radiative level shifts (Lamb shifts) require retaining modes in the range $a_0^{-1} < k < \lambda_c^{-1}$. For larger cut-offs $a_0^{-1} < k_F < \lambda_c^{-1}$, the material subsystems become more localised, but the individual interaction terms of the multipolar Hamiltonian cannot be considered weak. The {\em total} interaction can, however, be regarded as weak.

The regime of self-consistency of the electric dipole description has also been characterised via the dimensionless parameter $\mu = k_F R$. When $\mu \lesssim 10$, modes with $k \gtrsim a_0^{-1}$ begin to play an important role. In particular, for ordinary atomic systems the high densities required for the superradiant phase transition of the (dipolar) Dicke Hamiltonian are well within the regime $\mu \lesssim 10$. Indeed, as has been known for some time, the critical density occurs very near to the point at which atomic overlap effects begin to become important \cite{vukics_fundamental_2015}. However, our regularised and arbitrary gauge description does not reveal anything that would prohibit the phase transition as a matter of fundamental principle, such as a hitherto missed interaction term. In particular, engineered systems may offer a means by which to realise the required conditions while remaining within the regime $\mu \gtrsim 10$, for which even a dipolar description should be valid.\\

\section*{Appendix: Commutators and equations of motion of the local physical variables}\label{Appendix A}

We consider a single atom ($n=1$) with nucleus fixed at ${\bf 0}$. 

\subsection*{Commutation relations between local physical variables}

The non-vanishing commutators for the manifestly gauge-invariant local physical variables $\{{\bf r}\,,{\dot {\bf r}},\,{\bf E},\,{\bf B}\}$ are found using the definitions in Eqs. (\ref{Efield}) and (\ref{minimal coupling ansatz}), and ${\bf B}=\nabla\times {\bf A}$, along with the canonical commutation relations (\ref{com1}) and (\ref{com2}). They are
\begin{align}
    [r_i, \dot{r}_j]&= \frac{i}{m}\delta_{ij},\label{fundcomm1}\\
    [\dot{r}_i, \dot{r}_j]
    &= \frac{iq}{m^2} \epsilon_{ijk} B_{\varphi k}(\mathbf{r}), \label{fundcomm2}\\
    [\dot{r}_i, E_{j}(\mathbf{x})]
      &=\frac{iq}{m}\delta_{ij}\varphi(\mathbf{x}-\mathbf{r}), \label{fundcomm3}\\
    [E_{i}(\mathbf{x}), B_{j}(\mathbf{x'})]&=
-i\epsilon_{ijk}\nabla_{k} \delta(\mathbf{x}-\mathbf{x'}). \label{fundcomm4}
\end{align}
As required, each of these commutators is gauge ($g$)-independent. We provide the details in the case of Eq.~(\ref{fundcomm3}) below. The other commutators are obtained similarly.\\

\noindent We begin with
\begin{align}
m \dot r_i &= p_i - q A_{g\varphi,i}(\mathbf{r}),\label{rdp} \\
E_j(\mathbf{x}) &= -\Pi_j(\mathbf{x}) - P_{g\varphi,j}(\mathbf{x}),
\end{align}
therefore
\begin{align}
[\dot r_i, E_j(\mathbf{x})]
= -[\dot r_i, \Pi_j(\mathbf{x})] - [\dot r_i, P_{g\varphi,j}(\mathbf{x})].\label{rdE}
\end{align}
Using Eq.~(\ref{rdp}) together with
\begin{align}
A_{g\varphi,i}(\mathbf{r}) := \int d^3y\, \varphi(\mathbf{r}-\mathbf{y}) A_{g,i}(\mathbf{y}),
\end{align}
and
\begin{align}
A_{g,i}(\mathbf{y}) := A_{\rm T,i}({\bf y})+\partial^{\bf y}_i\int d^3 z\,  {\bm g}_{\rm T}({\bf z},{\bf y})\cdot  {\bf A}_{\rm T}(\mathbf{z}),
\end{align}
we obtain
\begin{align}
[\dot r_i, &\Pi_j(\mathbf{x})]
= -\frac{q}{m} [A_{g\varphi,i}(\mathbf{r}), \Pi_j(\mathbf{x})] \nonumber \\
&= -\frac{iq}{m} \int d^3 y\, \varphi(\mathbf{r}-\mathbf{y})
\left[
\delta^{\mathrm{T}}_{ij}(\mathbf{y}-\mathbf{x})
+ \partial_i^{\mathbf{y}} g_{\mathrm{T}j}(\mathbf{x},\mathbf{y})
\right].\label{Acont}
\end{align}
Using
\begin{align}
P_{g\varphi,j}(\mathbf{x})
:= -\int d^3 y\, g_j(\mathbf{x},\mathbf{y}) \rho_\varphi(\mathbf{y}),
\end{align}
and, from Eq.~(\ref{rdp}), the result
\begin{align}
[\dot r_i, \rho_\varphi(\mathbf{y})]
= -\frac{iq}{m} \partial_i^{\mathbf{r}} \varphi(\mathbf{y}-\mathbf{r})=\frac{iq}{m} \partial_i^{\mathbf{y}} \varphi(\mathbf{y}-\mathbf{r}),
\end{align}
we obtain
\begin{align}
[\dot r_i, P_{g\varphi,j}(\mathbf{x})]
&= -\frac{iq}{m} \int d^3 y\, g_j(\mathbf{x},\mathbf{y})
\partial_i^{\mathbf{y}} \varphi(\mathbf{y}-\mathbf{r}) \nonumber \\
&= \frac{iq}{m} \int d^3 y\,
[\partial_i^{\mathbf{y}} g_j(\mathbf{x},\mathbf{y})]\, \varphi(\mathbf{y}-\mathbf{r}).\label{Pcont}
\end{align}
The second equality follows from an integration by parts in which we have naturally assumed square integrability of ${\bf g}$ and $\varphi$ (vanishing at spatial infinity).

Combining the two contributions to Eq.~(\ref{rdE}) from Eqs.~(\ref{Acont}) and (\ref{Pcont}), we obtain
\begin{align}
[\dot r_i, E_j(\mathbf{x})]
=& \frac{iq}{m} \int d^3 y\, \varphi(\mathbf{y}-\mathbf{r})\nonumber 
\\&\times \left[
\delta^{\mathrm{T}}_{ij}(\mathbf{x}-\mathbf{y})+\partial_i^{\mathbf{y}} g_{{\rm T}j}(\mathbf{x},\mathbf{y})
- \partial_i^{\mathbf{y}} g_j(\mathbf{x},\mathbf{y})
\right].
\end{align}
Noting that ${\bm g} = {\bm g}_{\mathrm{T}} + {\bm g}_{\mathrm{L}}$, the contribution of ${\bm g}_{\rm T}$ is seen to vanish identically, ensuring gauge-invariance, and leaving
\begin{align}
&[\dot r_i, E_j(\mathbf{x})]\nonumber \\
&= \frac{iq}{m} \int d^3 y\, \varphi(\mathbf{y}-\mathbf{r})
\left[
\delta^{\mathrm{T}}_{ij}(\mathbf{x}-\mathbf{y})
- \partial_i^{\mathbf{y}} g_{\mathrm{L}j}(\mathbf{x},\mathbf{y})
\right].\label{semifin}
\end{align}
Since by definition
\begin{align}
\nabla\cdot \bm{g}(\mathbf{x},\mathbf{y}) \equiv \nabla\cdot \bm{g}_{\rm L}(\mathbf{x},\mathbf{y})= \delta(\mathbf{x}-\mathbf{y}),
\end{align}
the longitudinal component ${\bm g}_{\rm L}$ of ${\bm g}$ is determined uniquely as the gradient of the Green's function for the Laplacian:
\begin{align}
g_{\mathrm{L}j}(\mathbf{x},\mathbf{y})
= -\partial_j^{\mathbf{x}} \frac{1}{4\pi|\mathbf{x}-\mathbf{y}|}
= \partial_j^{\mathbf{y}} \frac{1}{4\pi|\mathbf{x}-\mathbf{y}|}.
\end{align}
It follows that
\begin{align}
-\partial_i^{\mathbf{y}} g_{\mathrm{L}j}(\mathbf{x},\mathbf{y})
= -\partial_i^{\mathbf{y}} \partial_j^{\mathbf{y}} \frac{1}{4\pi|\mathbf{x}-\mathbf{y}|}
= \delta^{\mathrm{L}}_{ij}(\mathbf{x}-\mathbf{y}),
\end{align}
and therefore
\begin{align}
\delta^{\mathrm{T}}_{ij}(\mathbf{x}-\mathbf{y})
- \partial_i^{\mathbf{y}} g_{\mathrm{L}j}(\mathbf{x},\mathbf{y})
= \delta_{ij} \delta(\mathbf{x}-\mathbf{y}).
\end{align}
Substitution of this result into Eq.~(\ref{semifin}) yields the final result in Eq.~(\ref{rdE}).

\subsection*{Equations of motion}

The equation of motion of observable $O$ is found in the gauge $g$ from
\begin{align}\label{vonneumann}
\dot{O}=-i[O,H_g].
\end{align}
In any gauge $g$ the Hamiltonian $H_g$ is the total energy given by Eq.~(\ref{en}). Using this expression, the commutators in Eqs.~(\ref{fundcomm1})-(\ref{fundcomm4}), and the operator identity $[A,B^2]= B[A,B]+[A,B]B$, we obtain both the Newton-Lorentz force law and the Maxwell-Ampere equation. We have
\begin{align}
    m\ddot {\bf r}&=-\frac{im^2}{2}[{\dot {\bf r}},{\dot {\bf r}}^2]-i{m\over 2}\int d^3 x[\dot {\bf r},{\bf E}(\mathbf{x})^2]\nonumber \\
    &=\frac{q}{2}\left[ \dot{\mathbf{r}}\times\mathbf{B}_\varphi(\mathbf{r})-\mathbf{B}_\varphi(\mathbf{r})\times \dot{\mathbf{r}}\right]+q\mathbf{E}_\varphi(\mathbf{r}),
\end{align}
and similarly
\begin{align}
     \dot{\mathbf{E}}(\mathbf{x})
     &=-\frac{im}{2}[\mathbf{E}(\mathbf{x}),\dot{\mathbf{r}}^{2}]-\frac{i}{2}\int d^3 x'[\mathbf{E}(\mathbf{x}),\mathbf{B}(\mathbf{x'})^{2}]\\
     &=-\mathbf{J}_{\varphi}(\mathbf{x})+\nabla \times \mathbf{B}(\mathbf{x}).
\end{align}

\bibliography{reg}

\end{document}